\documentclass [pre,floatfix,superscriptaddress,twocolumn]{revtex4-1}   	

\usepackage{graphicx}				
\usepackage{afterpage}	
\usepackage{amssymb}
\usepackage[version=3]{mhchem}
\usepackage{float}
\usepackage[flushleft]{threeparttable}

\usepackage{color}

\makeatletter
\newcommand{\colorcaption}[2][]{%
	\begingroup%
	\renewcommand{\@caption@fignum@sep}{ (color online). }%
	\caption[#1]{#2}%
	\endgroup%
}

\begin{document}			
	
	\title{Nonequilibrium limit cycle oscillators: fluctuations in hair bundle dynamics}
	
	\author{Janaki Sheth}
	\affiliation{Department of Physics and Astronomy, UCLA, Los Angeles California, 90095-1596, USA}
	
	\author{Sebastiaan W.F. Meenderink}
	\affiliation{ Auditory Research Center, Caruso Department of Otolaryngology, USC,  Los Angeles California, 90033, USA}
	
	\author{Patricia M. Qui{\~n}ones}
	\affiliation{Caruso Department of Otolaryngology, Keck School of Medicine, USC,  Los Angeles California, 90033, USA}
	
	\author{Dolores Bozovic}
	\affiliation{Department of Physics and Astronomy, UCLA, Los Angeles California, 90095-1596, USA}
	\affiliation{California NanoSystems Institute, UCLA, Los Angeles California, 90095-1596, USA}
	
	\author{Alex J. Levine}
	\affiliation{Department of Physics and Astronomy, UCLA, Los Angeles California, 90095-1596, USA}
	\affiliation{Department of Chemistry and Biochemistry, UCLA, Los Angeles California, 90095-1596, USA}
	\affiliation{Department of Biomathematics, UCLA, Los Angeles California, 90095-1596, USA}
	
	\begin{abstract}
		
		We develop a framework for the general interpretation of the stochastic dynamical system near a limit cycle.  
		Such quasi-periodic dynamics are commonly found in a variety of nonequilibrium systems, including the spontaneous 
		oscillations of hair cells of the inner ear.   We demonstrate quite generally that in the presence of noise, the phase of 
		the limit cycle oscillator will diffuse, while deviations in the directions locally orthogonal to that limit cycle will display the Lorentzian power 
		spectrum of a damped oscillator.  We identify two mechanisms by which these stochastic dynamics can acquire a complex frequency 
		dependence, and  discuss the deformation of the mean limit cycle as a function of temperature. The theoretical ideas are applied to data obtained from spontaneously oscillating hair cells of the amphibian sacculus. 
	\end{abstract}
	
	\pacs{XXX}

	\date{\today}	
	
	\maketitle
	
	\section{Introduction}
	
	Limit cycle oscillators~\cite{Strogatz1994} are ubiquitous in nonequilibrium systems, such as nonlinear electrical amplifiers~\cite{vanderPol1934}, 
	lasers~\cite{Simpson2014}, and chemical reactions~\cite{Field1973}. Within this last class of systems,  the Belousov-Zhabotinski reaction, 
	which generates  complex spatiotemporal patterns, has been particularly well studied~\cite{Zhabotinsky1991}.  Living systems are 
	replete with such nonlinear oscillators that control a wide  variety of temporal patterns~\cite{Winfree1967}, including circadian 
	rhythms~\cite{Biological-Clocks1960,Goldbeter1995,Mori1996,Goldbeter2002}, predator-prey dynamics~\cite{May1972}, 
	neuronal dynamics~\cite{Izhikevich2007,Schwab2010}, and cardiac rhythmogenesis~\cite{Glass1991}. The theoretical analyses of these 
	biological phenomena typically employ low-dimensional dynamical systems that are described by a small number of collective variables.  The
	relation of these collective variables to the plethora of underlying microscopic degrees of freedom is often poorly understood. Nonetheless, 
	these complex biological oscillators are often well modeled by such low-dimensional nonlinear systems.  
	
	The auditory system provides yet another example of such a nonlinear oscillator.  It performs extremely sensitive 
	mechanical detection, with the inner ear capable of detecting pressure waves that 
	result in {\AA}ngstrom-scale displacements \cite{Hudspeth2008}. While the biophysical mechanisms behind this sensitivity are not fully understood, a significant body of 
	experimental work indicates that an internal active mechanical process serves to amplify the incoming signal \cite{Hudspeth14, Robles2001}. The active process leads to an inherent 
	mechanical instability, causing the inner ear to generate sound in the absence of external input, in a phenomenon known as {\em spontaneous 
		otoacoustic emission} \cite{Manley2008}. Hence, the auditory system exhibits phenomena that indicate the presence of stochastically driven limit cycle oscillators. 
	While our analysis of these active systems will be general, we will demonstrate its application by analyzing the dynamics of the auditory system. 
	
	Detection in the inner ear is performed by mechanically sensitive hair cells, named for the bundles of stereocilia that protrude from their surfaces \cite{LeMasurier05}. 
	The stereocilia contain mechanically sensitive ion channels, which open and close as the hair bundle is deflected by sound \cite{Vollrath07}. This transduction 
	complex is connected to an internal array of molecular motors, composed of Myosin 1c, which continuously adjust the tension stored in the tip links connecting the 
	neighboring stereocilia ~\cite{Eatock2000}. The interplay between the opening and closing of the channels and the activity of the motors has been shown to lead to limit 
	cycle oscillations by the hair bundle \cite{Benser96, Martin2000}. Demonstrated \textit{in vitro} in a number of species, this active oscillation has been proposed to be one of the 
	mechanisms powering spontaneous emissions by the inner ear.
	
	Systems of nonlinear differential equations of varying levels of  complexity have been used to model the dynamics of active hair bundle motility, 
	as well as to describe the mechanical response of the full cochlea ~\cite{Reichenbach14, Martignoli2010, Martignoli2013, Gomez2014}. It was shown that the main characteristics of the auditory response can be 
	reproduced by a system that exhibits a supercritical Hopf bifurcation ~\cite{Camalet2000, Eguiluz2000, Stoop2003}. This suggests a relatively simple mathematical model of the 
	dynamics in terms of a two-dimensional, dynamical system.  The precise connection between these variables and the underlying hair cell structures is, however, 
	difficult to establish experimentally. As a consequence, a number of more complex models have been proposed, which are more directly 
	based on the known biophysical processes within the cell \cite{Martin03,Nadrowski2004,Han2010, Yuttana2011}. These models explicitly include terms related to the dynamics of the myosin motors, 
	the deflection of the stereocilia, and the electric potential across the cell membrane. The analysis of these more complex models reveals a rich 
	phase diagram, containing distinct dynamical phases separated by both continuous and discontinuous bifurcations ~\cite{Maoiléidigh2012}. Nonetheless, the 
	simple models of a supercritical Hopf bifurcation were shown to capture the fundamental hair cell dynamics that are of relevance to hearing ~\cite{Kanders2017, Salvi2015, Risler2004}. 
	
	One feature of the hair cell oscillators, and biological systems in general, is that all of the dynamic variables are subject to significant 
	amounts of stochastic noise ~\cite{Netten2003}. The deflection of the stereocilium, which occurs at the elastic pivot near its base, is subject to thermal Brownian 
	motion that leads to jitter of the bundle's position on the nanometer scale \cite{Nadrowski2004}. The forces exerted by the myosin motor complex 
	are subject to nonequilibrium, frequency-dependent noise associated with the attachment and detachment  of the motors, while they are climbing the actin network 
	that forms the internal core of the stereocillium \cite{Nadrowski2004}.  Secondly, opening and closing of transduction channels at finite temperature leads to 
	stochastic ``channel clatter'', which can lead to jitter in the bundle position~\cite{Dorval2006}.  Finally, the membrane potential 
	exhibits a stochastic driving term associated with both the channel clatter and the shot noise 
	in the ion transport across open ion channels~\cite{Lauger1975}.
	
	In this manuscript, we explore the effects of stochastic forces on two theoretical models of hair bundle dynamics, and compare these results to 
	experimental data obtained from hair bundles poised in the stably oscillating regime. Analysis of the response of a nonlinear system to noise is 
	typically performed by linearization about a stable fixed point, allowing one to write a system of linearized Langevin equations to describe the 
	stochastic dynamics. Hair bundles exhibiting spontaneous oscillations, however, are described by a limit cycle rather than a fixed point. 
	We propose that one may similarly linearize the system about a stable limit cycle. 
	
	We first consider a $d$-dimensional system of dynamical 
	variables $\vec{X}(t) = \left\{ x_{1}(t) , \ldots, x_{d}(t) \right\}$, obeying the noise-free (henceforth ``zero-temperature'') nonlinear 
	system of differential equations 
	\begin{equation}
	\label{basic-idea}
	\dot{\vec{X}} = \vec{F} \left( \vec{X} \right),
	\end{equation}
	where the dot denotes a time derivative. The function $\vec{F}$ depends on several parameters, whose values may be chosen to put
	the system in the limit-cycle regime. Thus, in the absence of noise, there exists a finite basin of attraction to a stable limit cycle solution of 
	Eq.~\ref{basic-idea}, with period $T$
	\begin{equation}
	\label{basic-idea-limit-cycle}
	\vec{X}_{0}(t) = \vec{X}_{0}(t + T),
	\end{equation}
	and which is nowhere stationary in time.  
	
	To analyze the effects of noise on the bundle dynamics, we linearize the system by introducing the Frenet frame associated with the 
	$d$-dimensional curve defining the zero-temperature limit cycle, Eq.~\ref{basic-idea-limit-cycle}. That orthonormal frame consists of one tangent vector $\hat{t}(s)$, one 
	normal vector $\hat{n}(s)$,  and $(d-2)$ other mutually orthogonal vectors $\hat{b}_{j}(s)$, $j =1, \ldots,  (d-2)$. In subsequent discussions, we analyse a three-dimensional system, wherein the frame comprises of $\{\hat{t}(s),\hat{n}(s), \hat{b}(s)\}$ and where $\hat{b}(s)$ is the binormal vector. All $d$ vectors of the Frenet frame 
	may be parameterized by a single independent variable $s$, which denotes the arclength along the limit cycle, measured from an arbitrarily 
	selected point on the cycle.   One may define a phase angle $\phi = 2 \pi s/\ell$, where $\ell$ is the arclength of the total limit cycle.  We will use $s$ and $\phi$ interchangeably
	in the following. 
	
	Using this parameterization, we will show that at finite temperature, fluctuations have a Lorentzian power spectrum in the $(d-1)$ 
	directions that are orthogonal to the local tangent of the zero-temperature limit cycle. The linear stability of the limit cycle forces these degrees of 
	freedom to behave effectively like overdamped oscillators in a thermal bath. The fluctuations in the tangent direction, however, generate 
	diffusive motion of $\vec{X}$ along the limit cycle. The first-order differential equations Eq.~\ref{basic-idea} provide no restoring force to 
	fluctuations that either advance or retard the motion of $\vec{X}$ in the tangent direction. 
	
	The effective potential for the Lorentzian variables and the effective diffusion constant of the arclength variable may 
	themselves be time dependent. There are two sources of this effect, which we call {\em mechanisms I and II}. Mechanism I is in effect when the 
	linearized equations of motion about the Frenet frame have an effective coupling between the Lorentzian variables and the arclength variable.  This 
	coupling modifies the diffusion constant of the phase variable in a frequency-dependent manner. Mechanism II occurs whenever the 
	effective potential for the Lorentzian variables, the zero-temperature speed of the phase point about the limit cycle, or the 
	effective diffusion constant of the arclength variable, are inherently arclength-dependent. As the system transverses its limit cycle, these variations make 
	the fluctuation spectrum vary in time.  Those variations generate extra structure in the power spectral density of the stochastic variables at discrete frequencies, 
	determined by the period : $\nu_{n} = 2 \pi n/T$, 
	where $n$ is an integer.
	
	The comparison between theory and the experimental measurements of hair bundle dynamics is fraught with one additional complexity.  The more complex dynamical
	models include at least three dynamical variables.  Previous experiments have measured a number of physiological parameters, including stereociliary position, membrane potential, calcium concentration, and others~\cite{Hudspeth2008},~\cite{LeMasurier05}, ~\cite{Vollrath07}. However, myosin motor activity during spontaneous oscillations is, so far, not directly observable.  In essence, only lower-dimensional projections of the full dynamical systems, are experimentally
	accessible. In this work, we study a two-dimensional projection of a 3-d system, with the projection plane defined by the position of the hair bundle and the membrane potential.  This raises a broader question: how can one use the measurement of a dynamical system to address the question of whether any ``hidden''
	dynamical variables are necessary or prevalent. 
	
	The remainder of the article is organized as follows. In section~\ref{sec:model-1}, we introduce and analyze a 
	two-dimensional model for a Hopf oscillator in the stably oscillating regime. In section~\ref{sec:model-2}, 
	we apply our analysis to a more complex three-dimensional model for hair bundle oscillations that explicitly includes the 
	experimentally hidden variable of motor activity. In section~\ref{sec:experiment}, we turn to experimentally observed noisy 
	spontaneous oscillations, exhibited by hair bundles of the amphibian sacculus. We find that predictions for the phase diffusion constant, based on {\em mechanism I} coupling of the simple supercritical Hopf system, are supported by the data.  We also observe some features indicative of 
	the proposed {\em mechanism II}.  Finally, we conclude in section~\ref{sec:summary}, where we review the relation between experiment and 
	theory regarding the fluctuation spectrum of these limit cycle oscillators, discuss future theoretical directions, and propose new experiments.

	\section{\label{sec:model-1}Two-dimensional dynamical system: Hopf oscillator}
	We begin with the simplest two-dimensional approach to the hair cell dynamics, the supercritical Hopf oscillator in its normal form~\cite{Strogatz1994}.  
	This is the lowest dimensional system that admits limit cycle oscillations. 
	The dynamical system can be described in terms of a single complex variable $z(t) = x(t) + iy(t)$ that obeys the (stochastic) differential equation
	\begin{equation}
	\label{Hopf-main}
	\dot{z} = z \left(  \mu - i \omega \right)  + b z |z|^2 + \eta_z(t).
	\end{equation}
	One may identify the real part $x(t)$ with the hair bundle displacement, and the imaginary part $y(t)$ with the internal active mechanism comprising of 
	elements such as myosin motor activity or calcium concentration. The details of these internal variables contributing to $y(t)$ 
	are not relevant for the proceeding discussion.   The zero-temperature dynamics of the model, {\em i.e.} when $\eta_{z}=0$, are controlled by the 
	values of the parameters  $\left\{\mu, \omega, b\right\}$.  We allow $b = b' + i b''$ to be a complex number whose real part is required to be positive 
	($b' > 0$) to ensure the stability of the limit cycle. When the (real) parameter $\mu < 0$, the system has a single stable fixed point at $z=0$ with 
	an infinite basin of attraction.  For $\mu >0$, this fixed point becomes unstable, and a circular limit cycle appears at radius 
	$R_{0} = \sqrt{\mu/b'}$.  The zero temperature  system traverses this circular limit cycle with a fixed angular velocity 
	$\omega_{0} = \omega + R_0^{2} b''$. 
	
	\begin{figure}
		\includegraphics[width=1\linewidth]{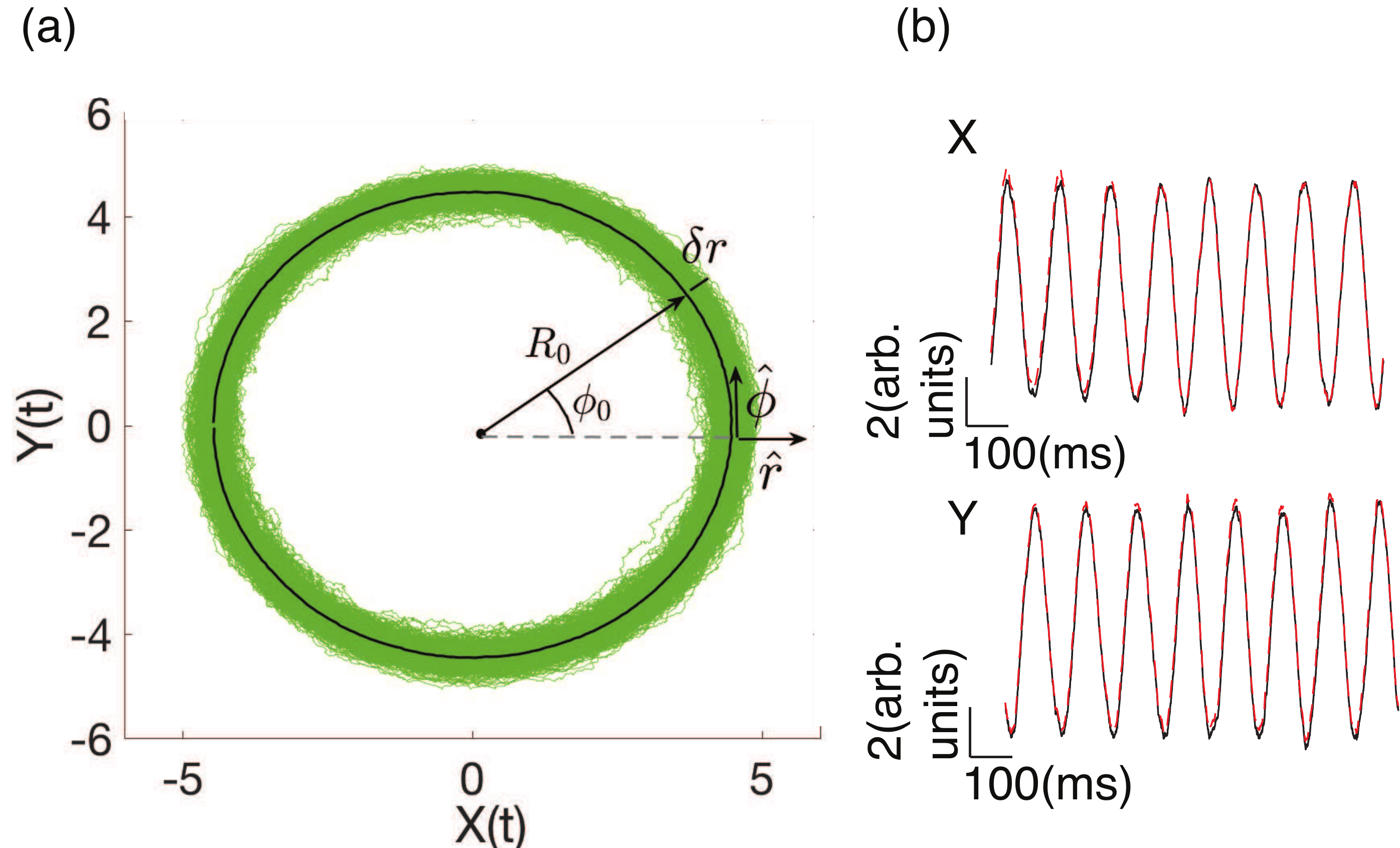}
		\colorcaption{{\bf Numerical simulation of the stochastic Hopf oscillator:}  Calculations were performed based on Eq.~\ref{Hopf-main}. (a) The finite-temperature (green) trajectories and 
			the mean (black) limit cycle. Also illustrated is the Frenet frame $\{ \hat{r},\hat{\phi} \}$ associated with the mean cycle. (b) A typical time series (black) of the stochastic dynamics of  $x(t)$ and $y(t)$. The 
			variables were low-pass filtered for further analysis (red dashed line).
			\label{fig:Limit-cycle-sim}}
	\end{figure}

	We assume that the stochastic forces acting on this system are Gaussian random variables with zero mean and a frequency-independent second moment.
	The latter point is not essential for our analysis; one may consider the effect of colored Gaussian noise with the same formalism.  
	In addition, we may assume that noise amplitudes along the orthogonal axes $x,y$ are statistically independent, but potentially selected from different Gaussian distributions.  Thus, we write
	\begin{eqnarray}
	\langle \eta_{i}(t) \rangle &=& 0\\
	\langle \eta_{i}(t) \eta_{j}(t') \rangle &=& A_{ij} \delta(t-t')
	\end{eqnarray}
	with $A_{ij}$ denoting elements of a diagonal matrix having two independent nonzero entries. The results obtained are consistent across several numerical values of $A_{ij}$. The range of values tested is further discussed in appendix A.
	
	Recasting Eq.~\ref{Hopf-main} in terms of polar coordinates 
	\begin{equation}
	\label{polar-form}
	z(t) = r(t)  e^{ i \phi(t) },
	\end{equation} 
	and then expanding about the zero-temperature limit cycle
	\begin{eqnarray}
	r(t) &=& R_{0} + \delta r (t) \\
	\dot{\phi}(t) &=& \omega_{0} + \delta \dot{\phi}(t)
	\end{eqnarray}
	we arrive at a local description of the fluctuations of the system in the limit cycle regime.  On substituting for $r(t), \phi(t)$ in Eq.~\ref{Hopf-main} and simplifying up to linear order, we derive
	\begin{eqnarray}
	\label{dr-dynamics}
	\delta \dot{r} &=& -2 \mu \, \delta r + \eta_r \\
	\delta \dot{\phi} &=& 2  b'' \sqrt{\frac{\mu}{b'}}\,  \delta r  +  \eta_{\phi}.
	\label{dphi-dynamics}
	\end{eqnarray}
	We have introduced the projections of the stochastic force onto the local normal $\hat{r}$ and tangent $\hat{\phi}$ to the 
	zero-temperature limit cycle, $\eta_{r}$ and  $\eta_{\phi}$ respectively.  It should be noted that this $\{\hat{r}, \hat{\phi} \}$ frame 
	is the Frenet frame for the circular limit cycle with $\hat{r} = \hat{n}$ and $\hat{\phi} = \hat{t}$.  
	\begin{figure}
		\centering
		\includegraphics[width= 1\linewidth]{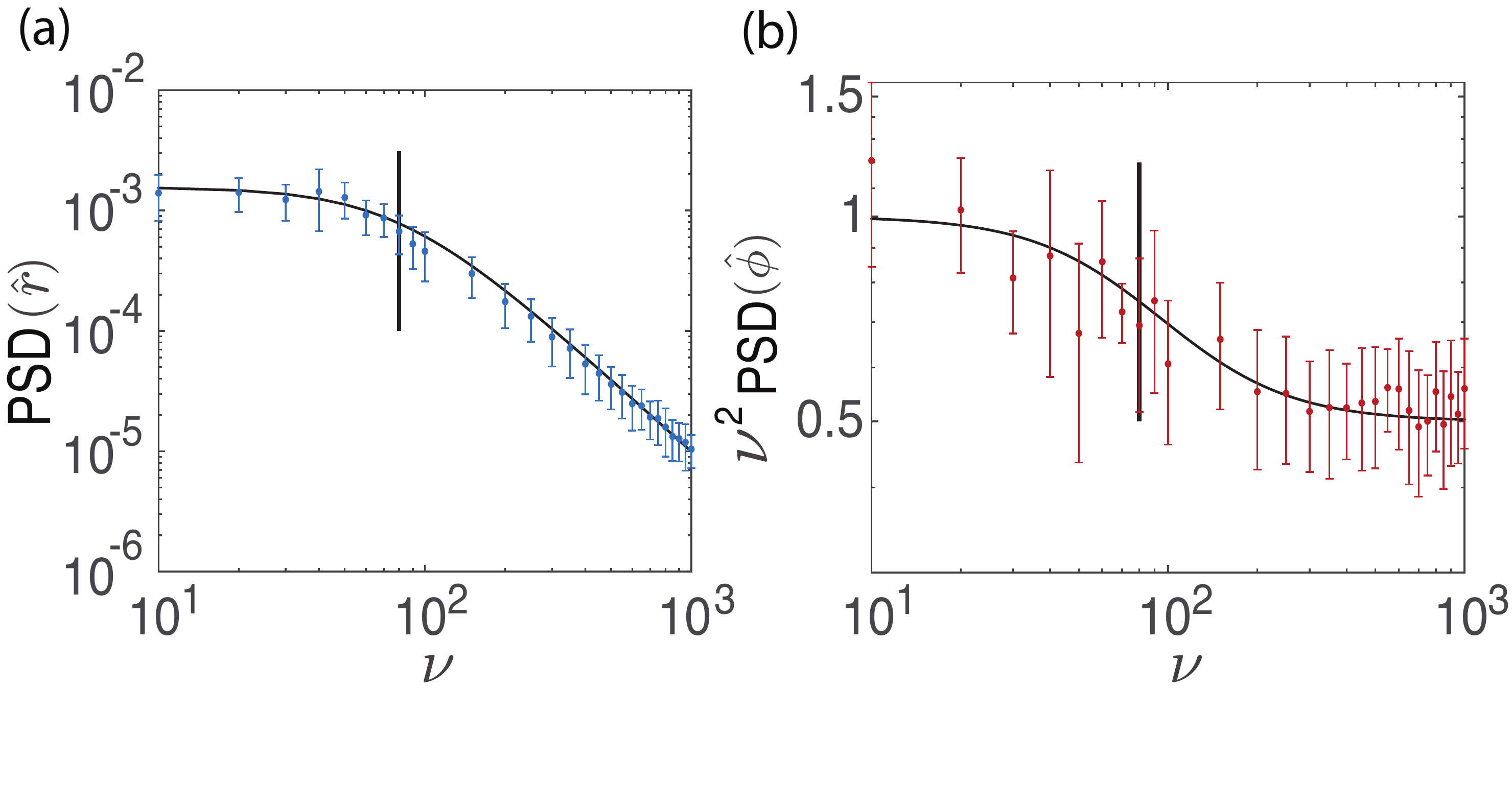}
		\colorcaption{{\bf Fluctuation spectra of the Hopf oscillator:} (a) Power spectral  density of fluctuations in the radial variable (blue dots) as a function of frequency $\nu$.  Error 
			bars denote standard deviations of the mean. The characteristic Lorentzian form has a corner frequency 
			$2 \mu$, marked by the vertical (black) line. (b) The frequency-dependent, effective phase diffusion constant, obtained from the
			product of the phase fluctuation power spectral density and $\nu^{2}$.  The spectrum exhibits a step, transition from a larger to a smaller 
			diffusion constant, at the corner frequency of the radial fluctuations (vertical black line). The 
			theoretical predictions - Eqs.~\ref{r-power-spectrum},~\ref{phi-power-spectrum} - are indicated with superposed black lines.
			\label{fig:PSDs-sim}}
	\end{figure}

	From  Eqs.~\ref{dr-dynamics},~\ref{dphi-dynamics}, we see that this two-dimensional system has one overdamped oscillator 
	(Lorentzian) degree of freedom, corresponding to displacements of the system normal to the zero-temperature limit cycle ($\delta r \hat{n}$), 
	and one diffusive arclength variable $s = R_0 \phi$.  In addition, we note that we should expect effects based on {\em mechanism I} to 
	generate extra frequency dependence in the diffusion about the limit cycle, arising from the coupling of $\delta r$ to the angular velocity variations 
	$\delta \dot{\phi}$, seen in Eq.~\ref{dphi-dynamics}.  Additionally, one observes that the non-zero parameter $b''$ controls the magnitude of this coupling, as it causes the fixed angular velocity of the system $\omega_0$ to be dependent on the radius.  We do not expect to observe effects of {\em mechanism II} in this system, 
	since the curvature of the potential for $\delta r$ is independent of $\phi$, as is the coupling between that variable and $\delta \dot{\phi}$.  
	Moreover, the mean velocity of the phase point of the zero-temperature dynamical system is independent of the arclength about the limit cycle. 
	
	If we make the further assumption that $\eta_{x}$ and $\eta_{y}$ are selected from the same ensemble, then the statistics of their projections 
	onto $\hat{r}, \hat{\phi}$ are equal and independent of arclength.  In that case, $A_{xx } = A_{yy} = A$, and we find the power 
	spectral density for radial fluctuations to be a simple Lorentzian:
	\begin{equation}
	\label{r-power-spectrum}
	\langle \left| \delta r(\nu) \right|^{2}\rangle  = \frac{A }{4 \mu^{2} + \nu^2}.
	\end{equation}
	The power spectrum of the phase fluctuations (our dimensionless arc length variable) is not simply diffusive, due to the mechanism I coupling between the
	normal and tangential fluctuations:
	\begin{equation}
	\label{phi-power-spectrum}
	\langle \left| \delta \phi(\nu) \right|^{2} \rangle  = A \left[ \frac{4 {b''}^{2} \mu}{b' \nu^2 \left(4 \mu^{2}+ \nu^2\right)} + \frac{ b'}{\mu \nu^2} \right].
	\end{equation}
	The effective diffusion constant of the phase variable is larger at low frequencies $\nu < 2 \mu$ than it is at higher ones. 
	The {\em mechanism I} coupling, in essence, adds extra phase noise from the overdamped fluctuations of the radial variable.  
	\begin{figure}
		\begin{center}
			\includegraphics[width=1\linewidth]{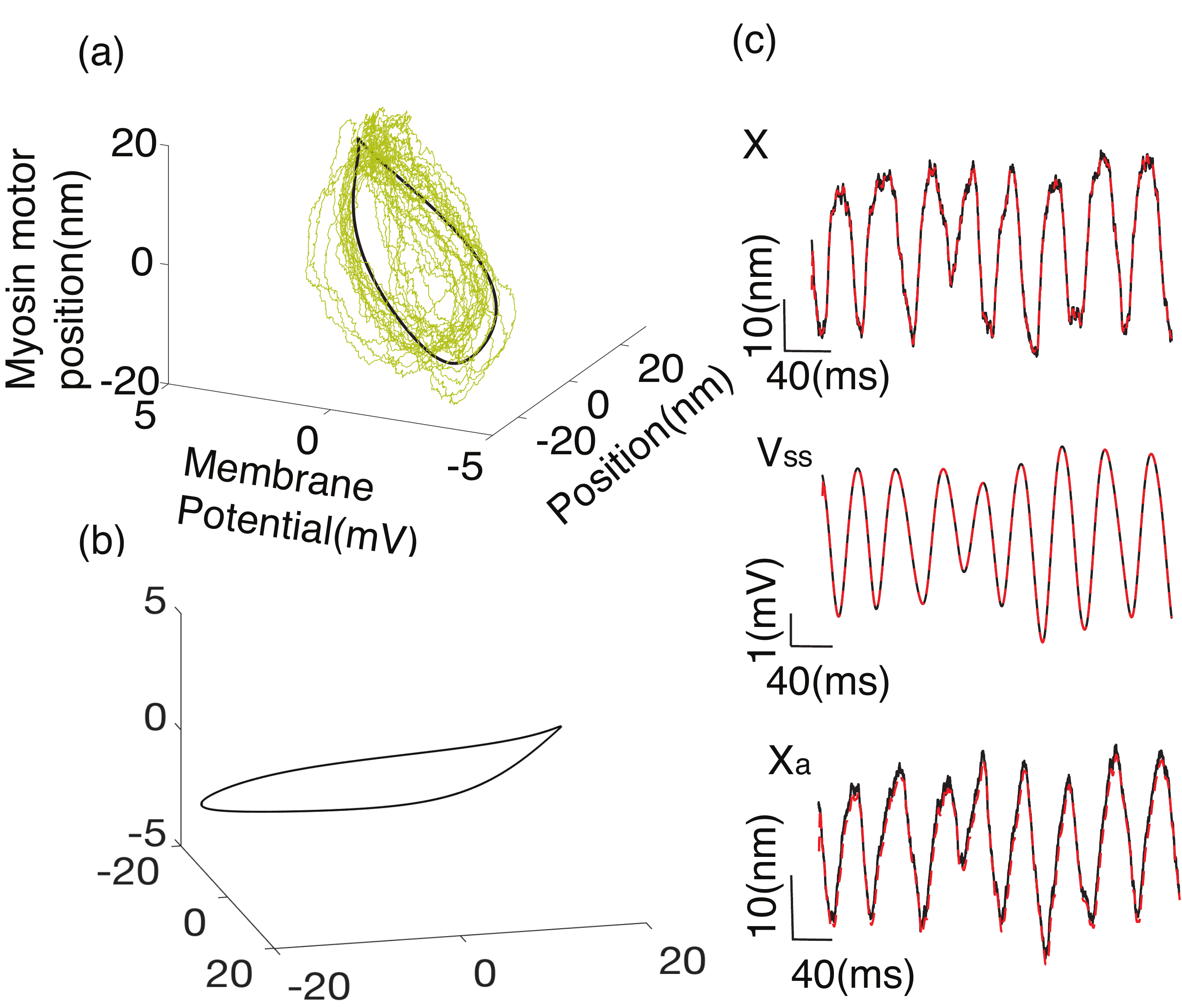}
		\end{center}
		\colorcaption{{\bf Numerical simulation of the three-dimensional hair bundle model:} Trajectories obtained by integrating Eqs.~\ref{Neiman-model-pos},~\ref{Neiman-model-myosin},~\ref{Neiman-model-V}. Left panels (a) and (b) illustrate two different perspectives of the nonplanar zero-temperature limit cycle (black). The green curve in (a) shows a representative trajectory about this cycle. Right panel shows typical time series (black) of the stochastic dynamics  of  the three variables and their respective lowpass filtered curves (red dashed). Similar to the Hopf simulation, the cutoff for the filter was chosen at $400\times2\pi$ Hz. The system noise temperature was chosen to be $T_{eff} = \tau_{noise} T = 0.25 T$. The constant $\tau_{noise}$ is indicative of the variance of the finite temperature hair bundle noise, and has value of 1 for a system obeying the fluctuation-dissipation theorem.
			\label{fig:Neiman-limit-cycle}}
	\end{figure}
	Since these radial fluctuations obey a Lorentzian power spectrum with a corner frequency of $2 \mu$, the effect of this cross-coupling 
	diminishes rapidly at frequencies higher than the corner frequency.  This effectively decreases the phase diffusion at higher frequencies. 
	
	We demonstrate these features of the fluctuation spectra using numerical simulations of the stochastic Hopf oscillator.  The simulation 
	details are standard and described in appendix A.   Fig.~\ref{fig:Limit-cycle-sim} illustrates the stochastic dynamics of the simulated supercritical 
	Hopf dynamical system.  The values of the Hopf parameters used to construct the same are: $\mu = 40$, $\omega_0$ = 10, $b'$ =  2, $b''$ = 2.
	Panel~\ref{fig:Limit-cycle-sim}(a) shows the  mean orbit of the stochastic system in the phase space spanned by $x,y$ (black line). The hair bundle phase space $\{ - \pi, \pi \}$ is partitioned into nearly 200 bins. Trajectories in each bin are then averaged, resulting in the mean limit cycle. For all forthcoming figures in the manuscript, similar methodology is applied to calculate the mean cycle. 
	
	Simulations provide us access to both 
	this averaged curve and the zero-temperature one, but in  experiments, we cannot access the latter. To better connect the simulations to the experimental observations 
	shown later in this manuscript, we study the behavior of these fluctuations about this averaged limit cycle. In this simple case, the average cycle is 
	similar to the zero-temperature limit cycle. We return to this point in our summary. 
	The green curves show a representative set of stochastic trajectories that meander about the mean limit cycle.  The local Frenet frame for the average system, 
	given by the unit vectors $\hat{r} = \hat{n}$ and $\hat{\phi} = \hat{t}$,  is indicated in the figure.  Panel~\ref{fig:Limit-cycle-sim}(b) shows typical time series of the variables $x,y$ after low-pass filtering. These were filtered at $400\times2\pi$ Hz; the Nyquist frequency was $500\times2\pi$ Hz.

	We numerically compute the power spectral density (PSD) of the stochastic deviations of the simulated trajectories about the mean limit cycle of the supercritical Hopf oscillator. The spectra of the fluctuations in the normal and tangential directions are shown in Figs.~\ref{fig:PSDs-sim}(a) and~\ref{fig:PSDs-sim}(b), respectively. In anticipation of the diffusive nature of the tangential
	or phase fluctuations, we plot the product of the tangential PSD  and $\nu^{2}$; note that this is the frequency-dependent phase diffusion constant.  To  obtain these results, 
	we project the state of the system onto the local Frenet frame, corresponding to the point of the mean limit cycle closest to the phase space point of the system. The perturbation of this state from the mean limit cycle, along the normal of the Frenet frame exhibits Lorenztian dynamics of an overdamped oscillator.  
	
	Fig.~\ref{fig:PSDs-sim}(b) is obtained from the drift-corrected difference between total arclengths traversed along the finite-temperature and the mean limit cycles. The arclength $(s)$ is given by hair bundle displacement as projected along the tangent of the local Frenet frame nearest to the phase space point. Thus, the total arclength for a bundle travelling along the mean limit cycle is equal to the cumulative lengths of all periods along the underlying zero-temperature curve. The presence of noise $\{\eta_r, \eta_{\phi}\}$ causes the finite-temperature arclength to differ from the mean curve arclength. The difference is corrected for the underlying drift, and the resulting frequency spectrum is given by Eq.~\ref{phi-power-spectrum}. The phase fluctuation $\delta \phi$ is equal to $2\pi (\delta s) /\ell$. Due to the coupling between the overdamped radial fluctuations and the phase velocity -- see Eq.~\ref{dphi-dynamics} -- the phase diffusion constant crosses over from
	a higher value, at frequencies below the corner frequency of the radial Lorentzian PSD, to 
	a lower one at higher frequencies; the crossover point of this step-like transition is marked by the vertical black line in both panels of Fig.~\ref{fig:PSDs-sim}.  The phenomenon has a simple interpretation.  Below the corner frequency, noise in the radial variable feeds back into the phase velocity 
	fluctuations.  Above the corner frequency, these radial fluctuations rapidly vanish, reducing the phase diffusion.    There are no {\em mechanism II} effects, since the mean
	phase velocity, the phase diffusion constant,  and the curvature of the effective potential in the radial direction are all phase independent.

	\section{\label{sec:model-2} Three-dimensional model of hair bundle dynamics}
	Hair bundle motility is more comprehensively described by higher-dimensional models that include multiple dynamical variables, aimed to accurately capture the 
	biophysical processes operant within the hair cell. We focus here on a particular version of the model, which includes two observable physiological variables  -- hair bundle displacement $X(t)$
	and the membrane potential of the hair cell $V_{ss}(t)$ -- as well as an internal variable $X_a(t)$ associated with the position of the 
	myosin motors along the actin filaments. For more details the reader is referred to Refs~\cite{Han2010, Neiman2011}. 
	This hair bundle model is defined by three nonlinear coupled differential equations:
	\begin{widetext}
		\begin{eqnarray}
		\label{Neiman-model-pos}
		\lambda\dot{X} &=& -K_{gs}\left (X - X_a -DP_0\right) - K_{sp}X + \eta_{x} \\
		\label{Neiman-model-myosin}
		\lambda_a\dot{X}_a &=& K_{gs}\left( X - X_a - DP_0\right) - \gamma F_{max} \left(1 - S_0 \left[1 + \alpha \frac{V_{ss}}{V_0} \right]P_0 \right) + \eta_{x_a}\\
		\label{Neiman-model-V}
		\ddot{V}_{ss} &=& -\omega_v \left( \frac{\beta_0}{\omega_v C_m} + \frac{g_t P_0}{\omega_v C_m} \right) \dot{V}_{ss} - \omega_v^2 
		\left(1 + \frac{g_t P_0}{\omega_v C_m}\right) \left(V_{ss} - V_0\right) - \frac{I_0 \omega_v}{C_m}.
		\end{eqnarray}
	\end{widetext}
	The model depends on sixteen physiological parameters, which are tabulated and described in appendix A.
	
	We refer to this model as {\em three-dimensional} since it relates three biologically relevant variables. As normally discussed in the theory of dynamical systems, this model exits in a four-dimensional phase space, since the differential equation governing the membrane potential is second order in time. As a consequence, we can show only a three-dimensional projection of the system's four-dimensional limit cycle.  We will see that the experimental data are confined to a two-dimensional projection of this three-dimensional limit cycle.
	
	Models based on the first two equations, without the cellular membrane potential, have been extensively studied. This system contains an actively driven mechanical 
	oscillator and is known to exhibit a dynamical phase portrait, exhibiting the so-called 'fish diagram'~\cite{Nadrowski2004}, with bifurcations separating quiescent, oscillatory, and 
	bi-stable states of the hair bundle.  These boundaries are controlled by both subcritical and supercritical Hopf bifurcations. The model employed here results from an extension of the two-dimensional model that includes the membrane potential (Eq.~\ref{Neiman-model-V}), described as an underdamped resonator with the same characteristic frequency as the bundle oscillator. 
	The electrical oscillator is bi-directionally coupled to the active mechanical one. Oscillations of the hair bundle affect the membrane potential via the mechano-electrical transduction (MET) channel current $g_t P_0$, leading to forward coupling. Here, $P_0$ denotes the opening probability of mechanoelectrical transduction channels and its equation is given in appendix A. Variations of the membrane potential modulate calcium influx, which affects the myosin motor activity; a dimensionless 
	parameter $\alpha$ controls the strength of this reverse coupling  ~\cite{Rami2014}.

	\subsection{Fluctuations around a three-dimensional zero-temperature limit cycle}
	We repeat our analysis of the fluctuations about the three-dimensional limit cycle. We introduce a third vector associated with the Frenet frame, the binormal vector defined by $\hat{b} = \hat{n} \times \hat{t}$.  We may now similarly resolve stochastic deviations of the system depicted in Fig.~\ref{fig:Neiman-limit-cycle}, along this mutually orthogonal triad of vectors, associated with each phase point on the zero-temperature limit cycle (black curve).
	
\begin{figure}
		\begin{center}	
			\includegraphics[width=1.05\linewidth]{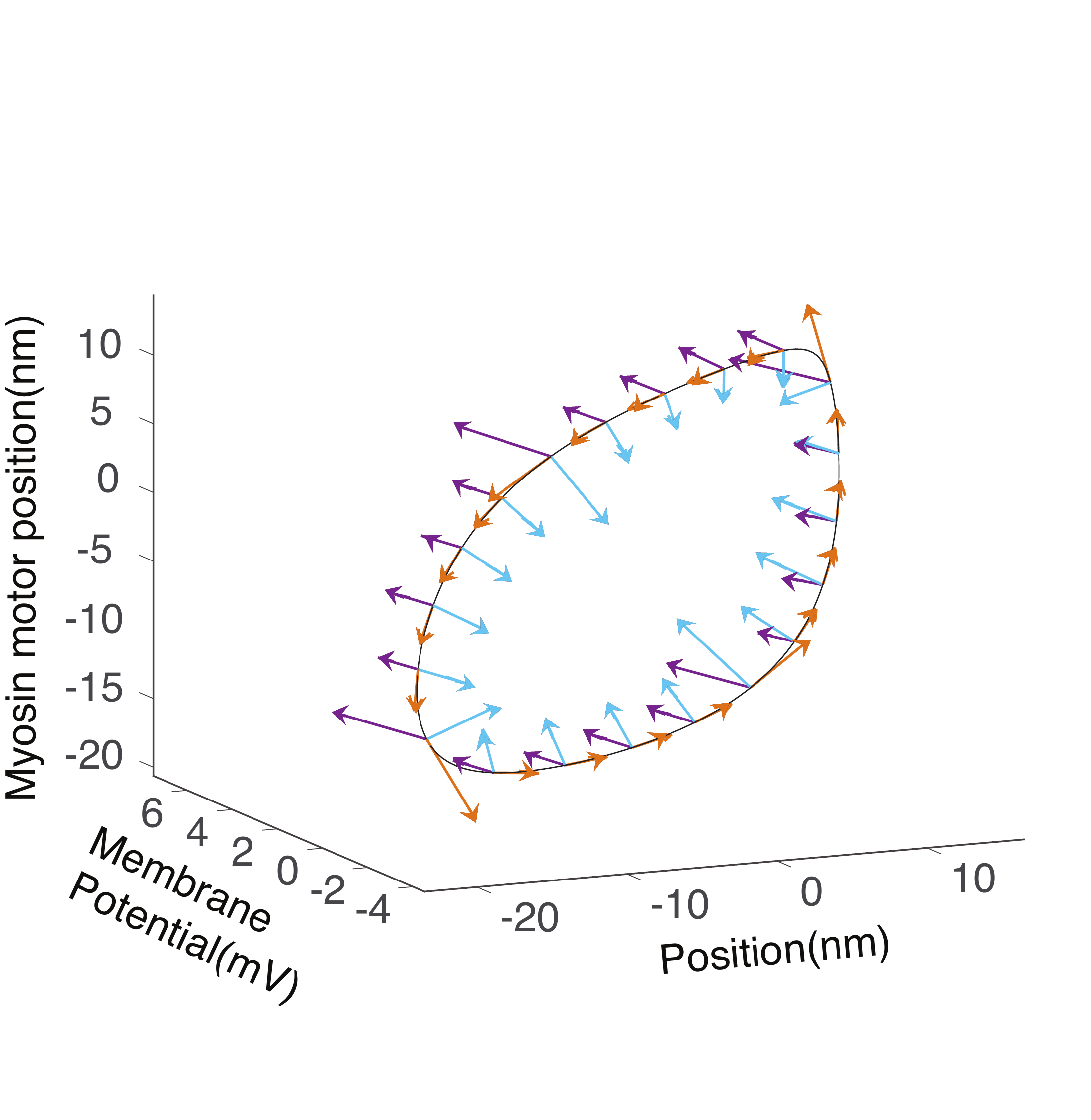}
		\end{center}
		\colorcaption{{\bf Rotating Frenet frame for the three-dimensional model:} The zero-temperature limit cycle is indicated with the orange line.
			The local Frenet frame $\left( \hat{t}, \hat{n}, \hat{b} \right)$ is 
			shown by the $\left(\mbox{orange}, \mbox{blue}, \mbox{purple} \right)$ unit vectors respectively, with the frames at four of the phase angles magnified by an arbitrary value for clarity. Note that the limit cycle is nonplanar.
			\label{fig:fernet-frame}}
	\end{figure}

	In our simulations, the variance of the noise in the stochastic trajectory (green curve) is modulated by the constant -- $\tau_{noise}$. Since the experimental system may be subject to nonequilibrium noise sources, we consider the effect on the hair bundle dynamics of a variable noise amplitude unrelated to the system's thermodynamic temperature. In effect we modify all noise amplitudes so that they reflect fluctuations at an effective noise temperature 
	$T_{eff} = \tau_{noise} T$. When $\tau_{noise} = 1$ the system obeys the fluctuation-dissipation theorem. Henceforth we refer to this specific value of $\tau_{noise} = 1$ as $\tau_0$.  In Fig.~\ref{fig:Neiman-limit-cycle}, the system noise temperature $T_{eff} = 0.25 T$.  The calculated three-dimensional zero-temperature limit cycle and the associated Frenet frames are shown in Fig.~\ref{fig:fernet-frame}.
	
\begin{figure}
		\includegraphics[width=1\linewidth]{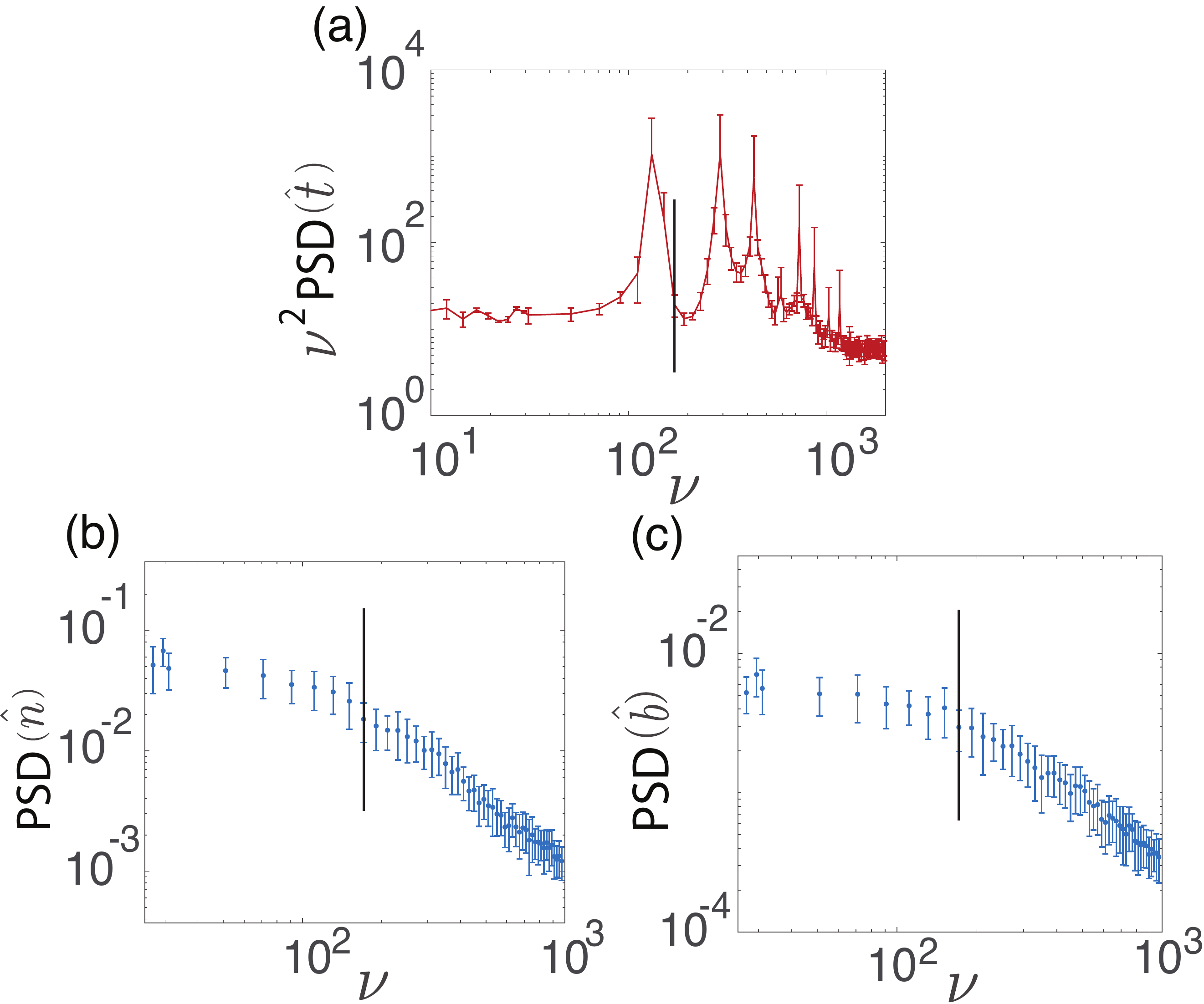}
		\begin{flushright}
			\colorcaption{{\bf Fluctuation power spectra for the three-dimensional model:} Fluctuations about the zero-temperature limit cycle were 
				projected onto the co-moving Frenet frame. The stochastic data use noise amplitude given by temperature $\tau_{noise} T= 0.25T.$ (a) The frequency-dependent diffusion constant. The spectrum shows peaks, as well as a crossover at the corner frequency (indicated by vertical black line), hinting to both mechanism I and II coupling. (b) The fluctuation power spectral density along $\hat{n}$. (c) The power spectral density of perturbations along $\hat{b}$. Both (b) and (c) exhibit Lorentzian spectra.
				\label{fig:psd-neiman-3d}}
		\end{flushright}
	\end{figure}

	The deviations of the stochastic system from its nearest point on the zero-temperature limit cycle are resolved along the unit vectors $\hat{t}$, $\hat{n}$, and $ \hat{b}$ of the local 
	Frenet frame to obtain the power spectral densities of their fluctuations in Fig.~\ref{fig:psd-neiman-3d}. We allude to the mean limit cycle case in later sections. Panels (b) and (c)
	show the (Lorentzian) relaxation of the degrees of freedom locally orthogonal to the limit cycle, in the normal (panel (b)) and binormal (panel (c)) directions. The fluctuations in these directions are those of overdamped oscillators. Panel (a) of this figure shows the frequency-dependent phase diffusion constant. Fig.~\ref{fig:psd-neiman-3d}(a) is obtained from the difference in total arclength of the zero and finite temperature oscillations, starting from an arbitrary phase point. The stochastic arclength variable is determined by the projection of the 
	distance traversed by the noisy trajectory onto the unit vector $\hat{t}$ associated with the Frenet frames of the deterministic limit cycle. Arclength ($s$) and phase ($\phi$) are interchangeable, and we depict fluctuations in the latter.
	
	This model exhibits new temporal structure in the phase diffusion constant due to effects of {\em mechanism II}. Both the mean phase velocity and the phase diffusion constant
	depend on the phase angle.  As a consequence, these quantities vary periodically in time as the system follows its limit cycle oscillations.  This periodic
	variation of the phase diffusion constant introduces structure in the diffusion constant at frequencies corresponding to the inverse limit cycle period $T$, {\em i.e.,} 
	at $\nu_{n} = 2 \pi n/T$, where $n = 1,2,\ldots$.  These features are analogous to the Bragg peaks associated with the Fourier transform of the spatial 
	density in a crystalline structure.   Superimposed on these peaks is the {\em mechanism I} effect observed in the simpler Hopf model.  There is a decrease in the effective diffusion constant for frequencies above the 
	corner frequency of the two Lorentzian degrees of freedom for the same reasons as discussed earlier. 
	
	\subsection{Lower dimensional projections}
	
	Realistic models of hair bundle dynamics include a number of variables describing the internal state of the hair cell, such as position of the myosin motors along the 
	stereocilia ($X_{\rm a}$), the forces they exert on the gating spring, internal calcium dynamics, and others \cite{Hudspeth14}. Currently, most of these variables are not accessible 
	experimentally. Typical recordings are limited to observations of  hair bundle mechanics, its oscillation and response to an imposed drive. Recently, 
	recordings of bundle mechanics were combined with electrophysiological recordings of the cell soma, in spontaneously oscillating and driven bundles \cite{Meenderink2015}. 
	This technique allows the simultaneous probing of two variables, the hair bundle position $X$ and the cell membrane potential $V_{\rm ss}$. 
	However, even with regard to the simplified model, discussed in prior section, this still allows access to only two of the three variables, as $X_{\rm a}$ remains ``hidden.''
	The experimentally accessible system is thus a projection of the full dynamical system onto a lower-dimensional manifold. 
	
\begin{figure}
		\begin{center}
			\includegraphics[width=1\linewidth]{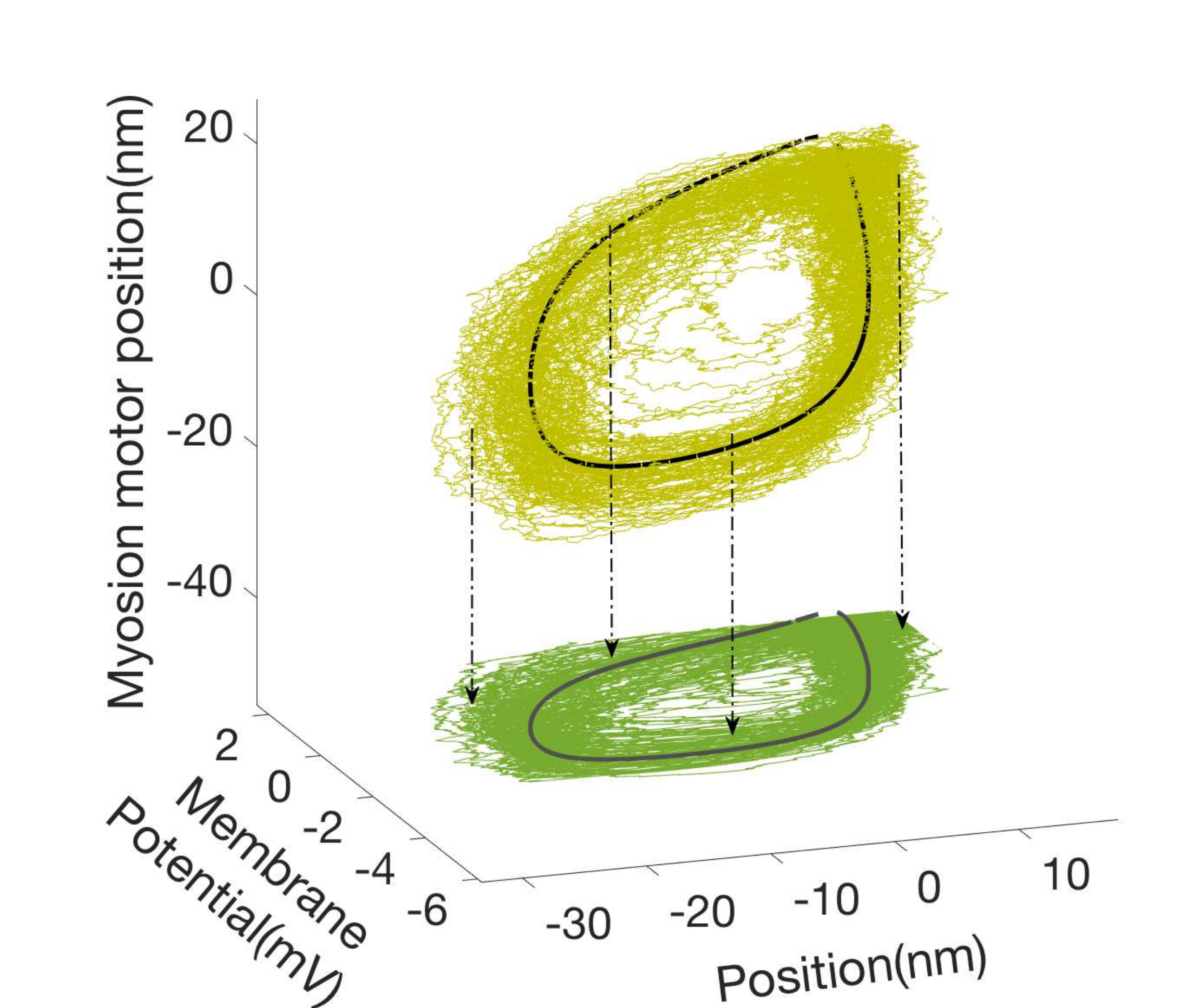}
		\end{center}
		\colorcaption{{\bf Projection onto the experimentally accessible plane:} The three-dimensional system contains only two experimentally accessible variables, spanning
			the ${\cal E}$ manifold. Stochastic trajectories (light green curves) about the deterministic three-dimensional limit cycle (black line) are observed only by their projection onto the ${\cal E} = (X,V_{\rm ss})$ plane. The projected trajectories (dark green) show fluctuations about a planar zero-temperature limit cycle (dark gray line). The inner two arrows denote the deterministic limit cycle projections, and the outer two point to the projected stochastic trajectory. To be compatible with 
			experimental data, $\tau_{\rm noise} / \tau_0 = 1$ for the given simulation, so $ <\eta_x>^2$ is $2kT\lambda$ and $<\eta_{x_a}>^2$ is $2kT\lambda_a$ .
			\label{fig:3d-proj-to-2d}}
	\end{figure}

	In our system, we have experimental access (discussed in a subsequent section) to the two dimensional ${\cal E} = (X,V_{\rm ss})$ plane. To examine the implications of this projection using our numerical model, we start with the three-dimensional noisy limit cycle, and 
	project it onto the experimental manifold ${\cal E}$, as shown in Fig.~\ref{fig:3d-proj-to-2d}. The full three-dimensional limit cycle is 
	shown in light green, and the two-dimensional projection of the numerical simulation is shown in dark green. These curves have been 
	simulated using noise amplitudes determined by Fluctuation-Dissipation theorem at room temperature ~\cite{Kubo1966}. The figure also depicts the zero-temperature limit cycle for the three- and two-dimensional systems, in black and gray, respectively.  From here on, to ensure 
	consistent comparisons with experimental data, we analyze the simulated data in this two-dimensional projection.

	\subsection{Mean limit cycle}
	
	\begin{figure}
		\includegraphics[width=1\linewidth]{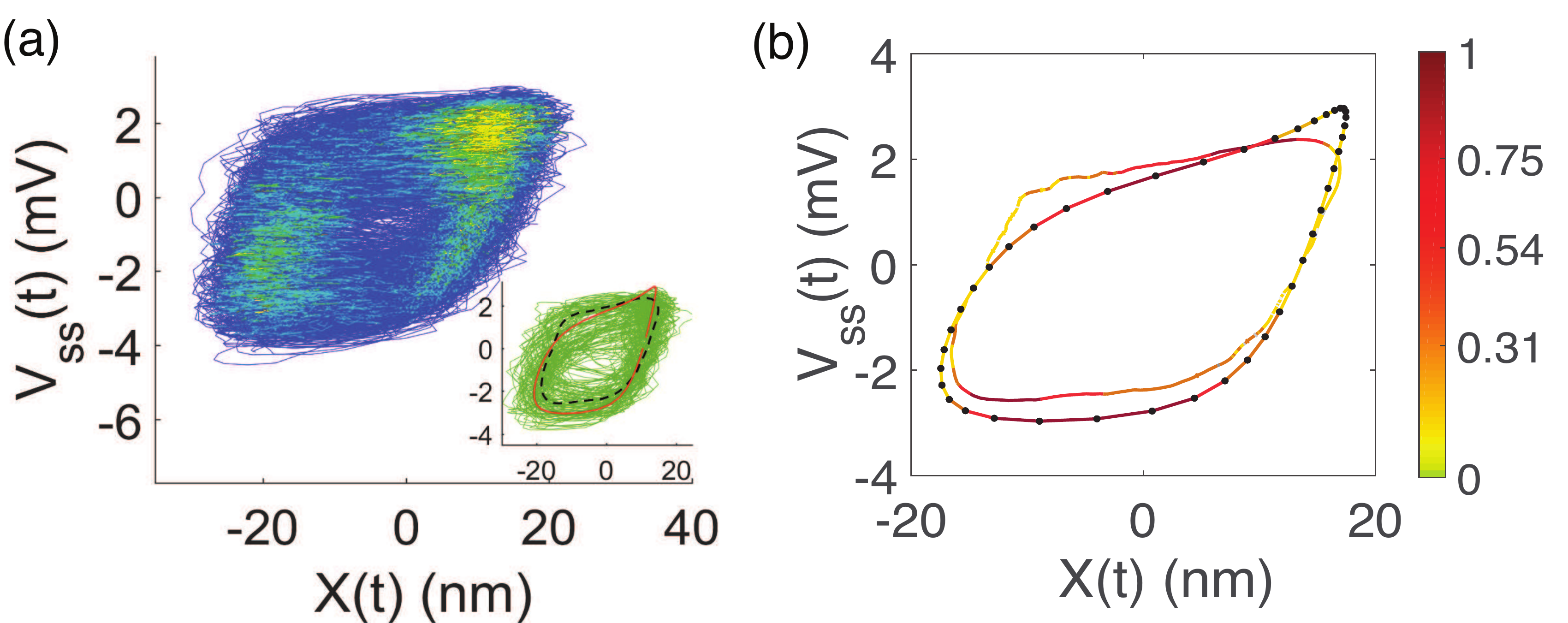}
		\colorcaption{{\bf Effects of finite temperature:} (a) A representative stochastic trajectory of length $t = 1060 T$, where T is the mean period, is depicted using a density plot. The less dense regions are in dark blue. Yellow indicates those with the highest density. (Inset) The zero-temperature (red) and mean (dashed black) limit cycle, computed at finite-temperature $T_{eff} = \tau_{0} T = T$, are superposed on the stochastic trajectory (green).(b) The mean two-dimensional limit 
			cycle in the ${\cal E}$ manifold, for both the deterministic (connected dots) and stochastic (line) systems. The mean 
			velocity of the phase point is denoted by a color map, where yellow (light gray)) color corresponds to lower and red (dark gray) to higher phase velocity. 
			The velocities, as indicated on the color bar, are normalized to the maximum velocity of the hair bundle along its mean cycle, and are hence dimensionless.
			\label{fig:velocity-of-limitcycle}}
	\end{figure}
	The analysis of experimental data introduces another complexity, briefly alluded to in a prior section. The biological problem does not 
	provide access to the noise-free system.  While the two were nearly identical in the case of the supercritical Hopf oscillator, in a more complex system, 
	the observable mean limit cycle may be different from the zero-temperature one.  In fact, both the size and the shape of the limit cycle may change with the noise 
	amplitude, in an effect that is analogous to the thermal expansion of crystalline materials. The mean 
	limit cycle of the dynamical system can distort due to the fact that the potential describing the stable limit cycle typically also has anharmonic terms. 
	
	We illustrate the noise dependence of the shape of the mean limit cycle (dashed black) in the experimentally accessible manifold ${\cal E}$ depicted in the inset of  Fig.~\ref{fig:velocity-of-limitcycle}. 
	In panel (a), we represent the finite-temperature limit cycle using a density plot, where the regions get denser as the colors traverse from dark blue to light blue to yellow. This is indicative of the phase-dependent properties of the limit cycle. In the inset, we show the deterministic limit cycle (red) and the superposed mean limit cycle (dashed black), with finite noise in all three dynamical variables.  As shown in the figure, the mean limit cycle distorts at finite temperature. The noise amplitudes 
	were chosen to represent equilibrium fluctuations at room temperature, as determined by the fluctuation-dissipation theorem.  We note that this choice constitutes a simplification, 
	particularly with respect to the variable reflecting myosin motor activity, where one might reasonably expect both colored noise and noise amplitudes unrelated to the 
	dissipative terms in the equation of motion. We will explore these effects in a subsequent study~\cite{Janaki-unpublished}.
	
	We observe that the zero-temperature limit cycle generally has sharper features -- smaller radii of curvature -- than the noisy one, as is evident, for example, in the upper right corner of the inset in panel (a). This is also noticed in the upper right corner of the limit cycles in panel (b).We suggest that this rounding of the zero-temperature limit cycle by noise is a generic feature, as stochastically driven excitations away from the zero-temperature limit cycle allow for cutting of any sharp corners through an activated escape mechanism.  At such sharp corners, the local basins of attraction of the two parts of the limit cycle necessarily become close so that the system may thermally jump from one branch to the next.  In particular, sharp corners with slow phase speeds allow for both a lower barrier and a greater number of statistically independent attempts at barrier crossing, so that these features show the strongest
	rounding of the corners. Indeed, we see this from a plot of  the mean velocity of the phase points on the limit cycle, illustrated as a heat map 
	superposed on the zero-temperature limit cycle oscillation in panel (b) of Fig.~\ref{fig:velocity-of-limitcycle}, where cooler (yellower) colors denote slower speeds. We also notice that the slower speeds correspond to the denser regions in panel (a).
	An analysis of such barrier crossing events using Kramers escape theory is forthcoming~\cite{Janaki-unpublished}.
	
	\begin{figure}
		\includegraphics[width=1\linewidth]{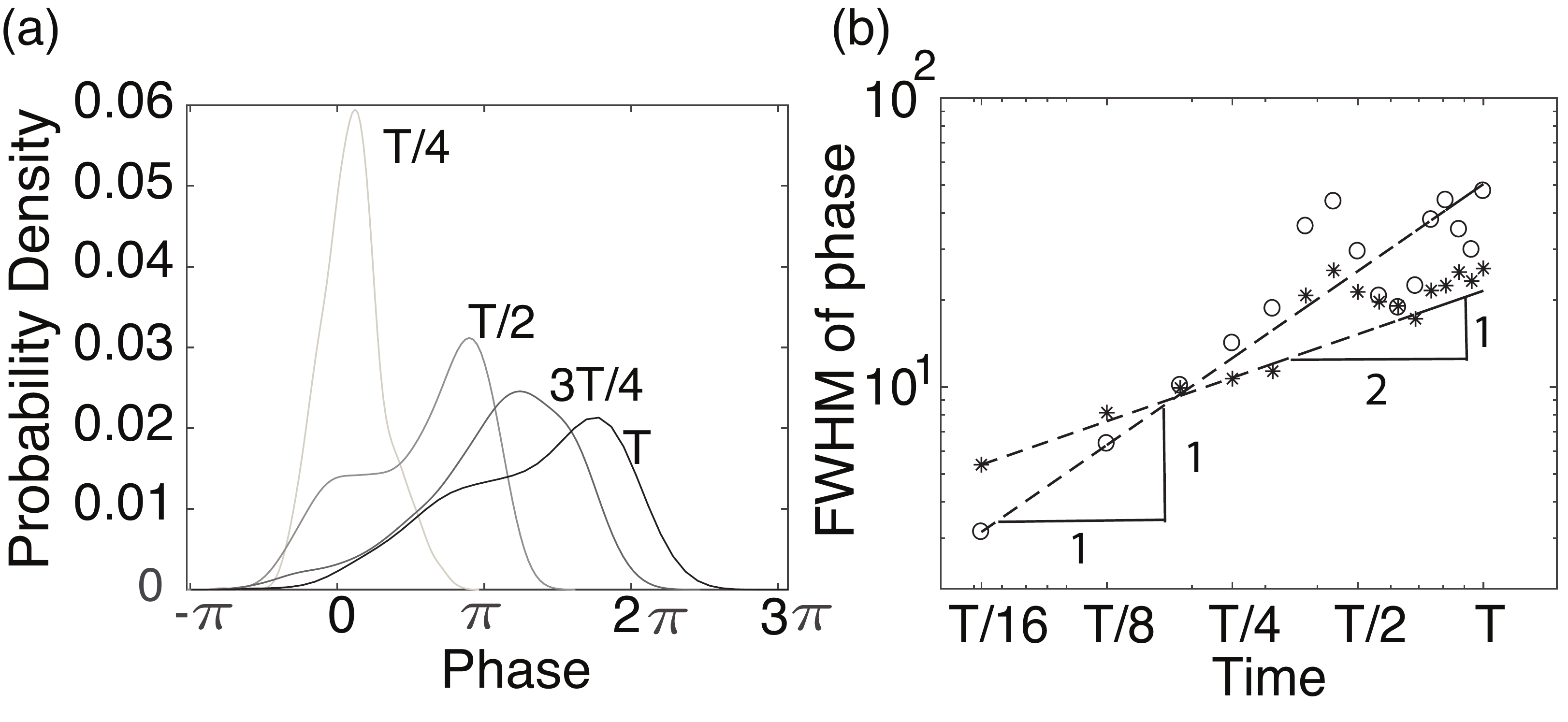}
		\caption{{\bf Phase diffusion along the limit cycle:} (a) The time evolution of an ensemble of systems, synchronized at a fixed but arbitrary initial phase point.
			(b) The Full Width Half Maximum (FWHM) of the phase distribution is not simply proportional to $\sqrt{t}$. The points represent two data sets, evaluated at time delays of ($T/16, T/8$ ... $15T/16, T$), starting from two different initial phase points (shown as stars and open circles). We observe advective and diffusive spreading of the ensembles, as seen by the two slopes of 1/2 and 1.
			\label{fig:phase-distribution}}
	\end{figure}
	
	We explore the effects of phase diffusion in another way.
	At time $t=0$, an ensemble of systems is poised at an arbitrary but fixed phase point on the limit cycle. Each system evolves in time under the action of the stochastic differential equations, Eqs.\ref{Neiman-model-pos},\ref{Neiman-model-myosin},\ref{Neiman-model-V}, with noise variances of $2kT \times \tau_{\rm noise} / \tau_{0} = 2kT$ .
	We compute the distribution of their phases (reported in terms of the total phase traversed about the limit cycle) at time delays corresponding to $T/4,T/2,3T/4,T$, 
	where $T$ is the mean period of the stochastic limit cycle. As shown in panel (a) of Fig.~\ref{fig:phase-distribution}, the distribution of phase points exhibits an 
	asymmetric spreading.  We note that the spreading of the phase distribution is not simply proportional to $\sqrt{t}$, as can be seen in Fig.~\ref{fig:phase-distribution}(b), 
	which is to be expected from the frequency-dependent phase diffusion constant. This spreading is also impacted by advection, as the ensemble of stochastic systems 
	converge or diverge in the lower and higher velocity regions, respectively. The dominant effect that is observed depends upon the sequence of regions that the systems 
	encounter. The open circle ensemble is dominated by advective spreading due to the local change in mean phase velocity along the limit cycle, while the star one
	is dominated by diffusive spreading. Moreover, we see that the distribution is not a simple Gaussian, as would be expected from a normal 
	advection-diffusion equation.  The variation of the mean phase velocity with phase accounts for most of this effect. Finally, we note that, after only one 
	period $t=T$, the width of the phase distribution is comparable to the total phase around the limit cycle $(2 \pi)$.  Hence, for this level of noise, the 
	ensemble that was phase synchronized at $t=0$ is (nonuniformly) distributed around the entire limit cycle after only one period.
	\begin{figure}
		\includegraphics[width=1.00\linewidth]{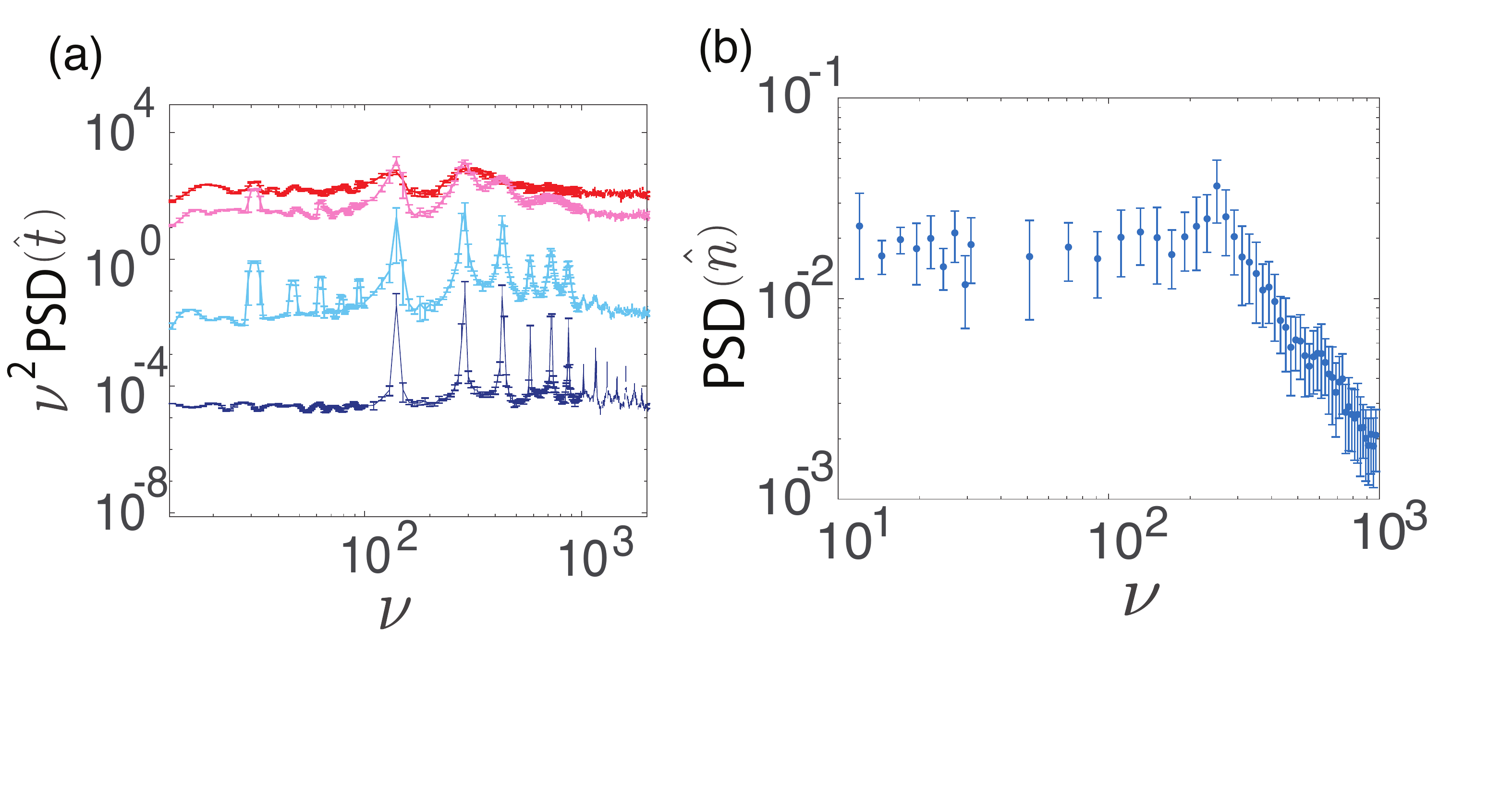}
		\begin{flushright}
			\colorcaption{{\bf Fluctuation spectra of the projected system:} (a) Frequency-dependent phase diffusion constant $D_{\rm s}$, obtained at four different noise variances: 
				$\tau_{\rm noise}/\tau_{0} \times 2kT = 0.005 kT, 0.05 kT, 0.5 kT, 2 kT$ (blue (lowest), cyan (second from bottom), pink (second from top), red (topmost)). For $\tau_{\rm noise}/\tau_{0} =0.0025$, the system is nearly deterministic. 
				Increasing noise amplitude broadens the Bragg peaks, as the system loses phase coherence over a time $T^{\star} \sim \ell^{2}/D$.  
				The spectra have been shifted vertically for visibility by multiplying by $( 10^{-4}, 10^{-2}, 1, 1)$.  
				(b) In the two-dimensional projection comprising of X,V$_{ss}$, the orthogonal direction is $\hat{n}$. The power spectrum of fluctuations along this direction is Lorenztian.
				\label{fig:temp-dependence-of-D}}
		\end{flushright}
	\end{figure}
	
	A simple criterion can be developed to account for the emergence of phase coherent effects, such as the peaks arising from {\em mechanism II} effects.  
	Questions regarding phase coherence can be recast into a statement about the ratio of the phase diffusion time $T^{\star} = \ell^{2}/D$ to the mean period $T$.  Herein, $\ell$ refers to the total arclength, measured in units of limit cycle periods, over which the system is coherent. 
	When $T^{\star} \gg T$, the period of phase coherence is equal to many limit cycle periods, and this phase coherence will 
	result in the appearance of peaks due to {\em mechanism II} effects.  On the other hand, for sufficiently short phase coherence times $T^{\star} \ll T$, any periodic structure of the deterministic limit cycle oscillator is lost, 
	and the peaks disappear from the fluctuation spectrum.  
	
	We observe this transition in our numerical simulations by adjusting the amplitudes of the noise, while holding the other parameters fixed.  The results are shown 
	in Fig.~\ref{fig:temp-dependence-of-D}, where we produced numerical data based on the Eqs. \ref{Neiman-model-pos},~\ref{Neiman-model-myosin},~\ref{Neiman-model-V} and projected the results onto ${\cal E}$.  The resulting spectra of the
	phase diffusion constant, obtained at four different noise levels, are shown in the panel (a).  The dark blue curve measures phase diffusion for an effective noise 
	variance of $\tau_{\rm noise}/\tau_{0} \times 2kT = 0.005kT$.  In this nearly deterministic system, phase coherence, as defined above, is maintained for $\sim 40$ periods.  As a consequence, we observe a full sequence of Bragg peaks.  Upon increasing the noise variance to $\tau_{\rm noise}/\tau_{0} \times 2kT = 0.05kT$ (light blue), we see that 
	these peaks weaken and broaden. The phase coherence lasts for $\sim 30$ periods.  Only the principal and second peak is observable at $\tau_{\rm noise}/\tau_{0} \times 2kT = 0.5kT$ (light pink); these are barely visible at $\tau_{\rm noise}/\tau_{0} \times 2kT= 2kT$ (red), as expected from the previous analysis. The time scales over which the system 
	loses phase coherence are equal to $\sim 5$ and $\sim 1$ periods, respectively.
	
	\begin{figure}
		\includegraphics[width=1\linewidth]{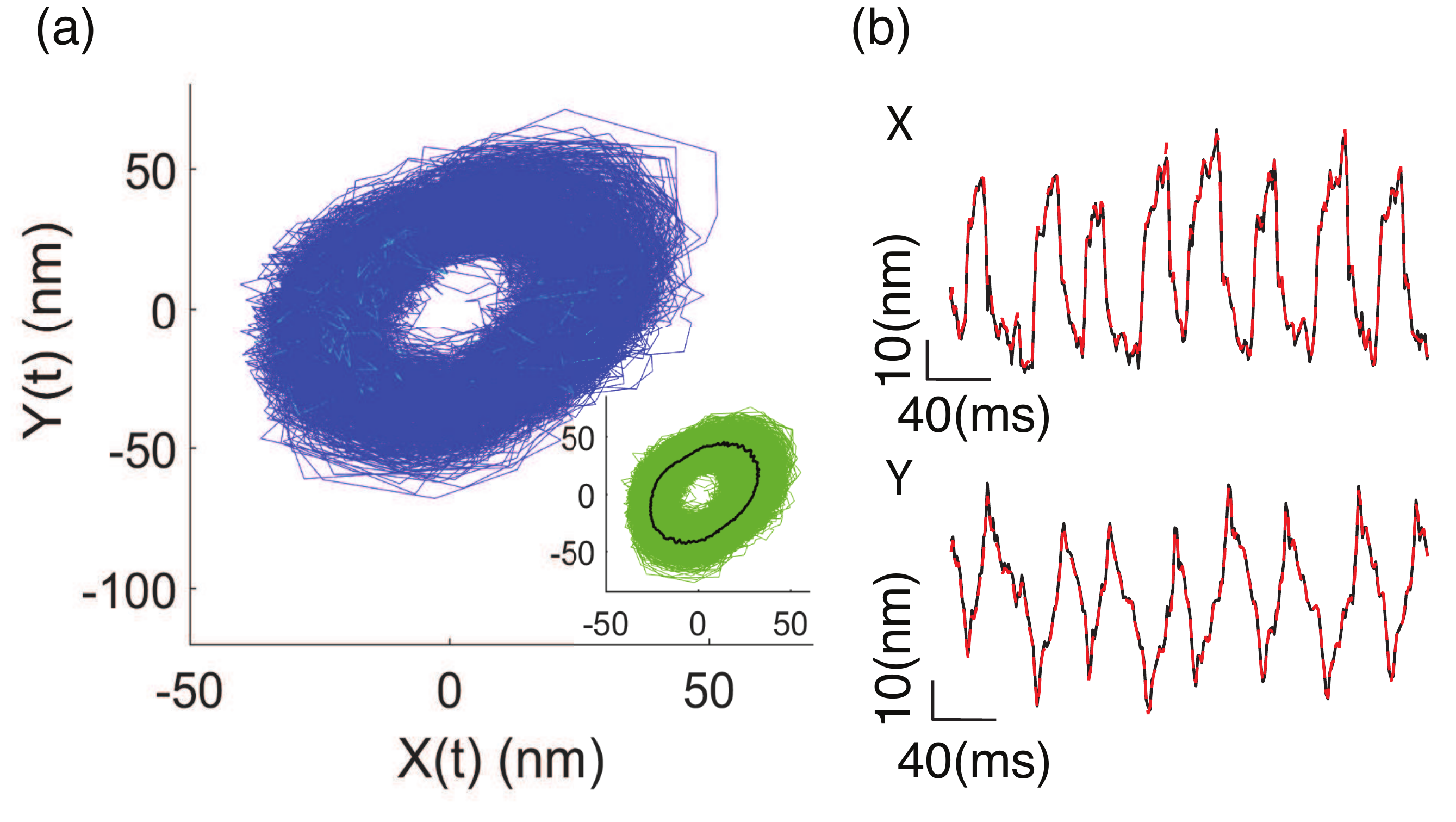}
		\colorcaption{{\bf Experimental recordings of spontaneous bundle oscillations:} (a) Density plot of a typical limit cycle trajectory (blue curve) 
			of a saccular hair bundle in the phase space spanned by bundle position $X(t)$ and its Hilbert transform $Y(t)$. (Inset) Additionally, the superposed averaged limit cycle is shown in black. (b) Experimental recordings of hair bundle position $X(t)$ and its Hilbert transform $Y(t)$ (black curves); the data were corrected for slow drift 
			innate in these biological preparations. The red dashed line show the low-pass filter time series used to construct the limit cycle.
			\label{fig:limit-cycle-expt1}}
	\end{figure}
	
	\section{\label{sec:experiment}Experimental observations}
	To test the theoretical predictions from the prior sections, we compare them to experimental observations of hair bundle dynamics. Recordings 
	were obtained from \textit{in vitro} preparations of the amphibian sacculus, following techniques described in earlier publications~\cite{Ramunno2009}. Biological preparations were mounted in chambers that allow optical access to hair bundles, while maintaining their active process~\cite{Martin2001,Martin1999,Benser1996}. Time traces of spontaneous hair bundle oscillations were obtained 
	from twenty cells, exhibiting a broad range of limit cycle frequencies and amplitudes, as well as variation in the amplitude of the 
	fluctuations about the mean limit cycle. Hair cells were pre-selected that exhibited only one mode of oscillation, hence a single peak in the 
	spectral density; cells that showed more complex multi-mode oscillation were not considered in the current study. The experimental data depicted in Fig.~\ref{fig:limit-cycle-expt1}, ~\ref{fig:limit-cycle-3d-expt1} were low-pass filtered at $400\times2\pi$ Hz (red-dashed); the Nyquist frequency was $500\times2\pi$ Hz. 
	
	\subsection{Comparison to the supercritical Hopf system}
	The Hopf variables are related to each other by the Hilbert transform, which has  previously been used to create the two-dimensional phase space for experimental data \cite{Salvi2016}. We follow that procedure, and in panel (b) of Fig.~\ref{fig:limit-cycle-expt1}, show the time traces of the hair bundle position and its Hilbert transform. The panel (a) of Fig.~\ref{fig:limit-cycle-expt1} displays the density plot of the experimentally observed noisy limit cycle (blue). The superposed mean limit cycle (black) is shown in the inset. One observes the uniformity of phase space density which points to the lack of {\em mechanism II}. A hair bundle that exhibited more irregular oscillation is shown in Fig.~\ref{fig:limit-cycle-expt2} in appendix B. 
	
\begin{figure}
		\includegraphics[width=1\linewidth]{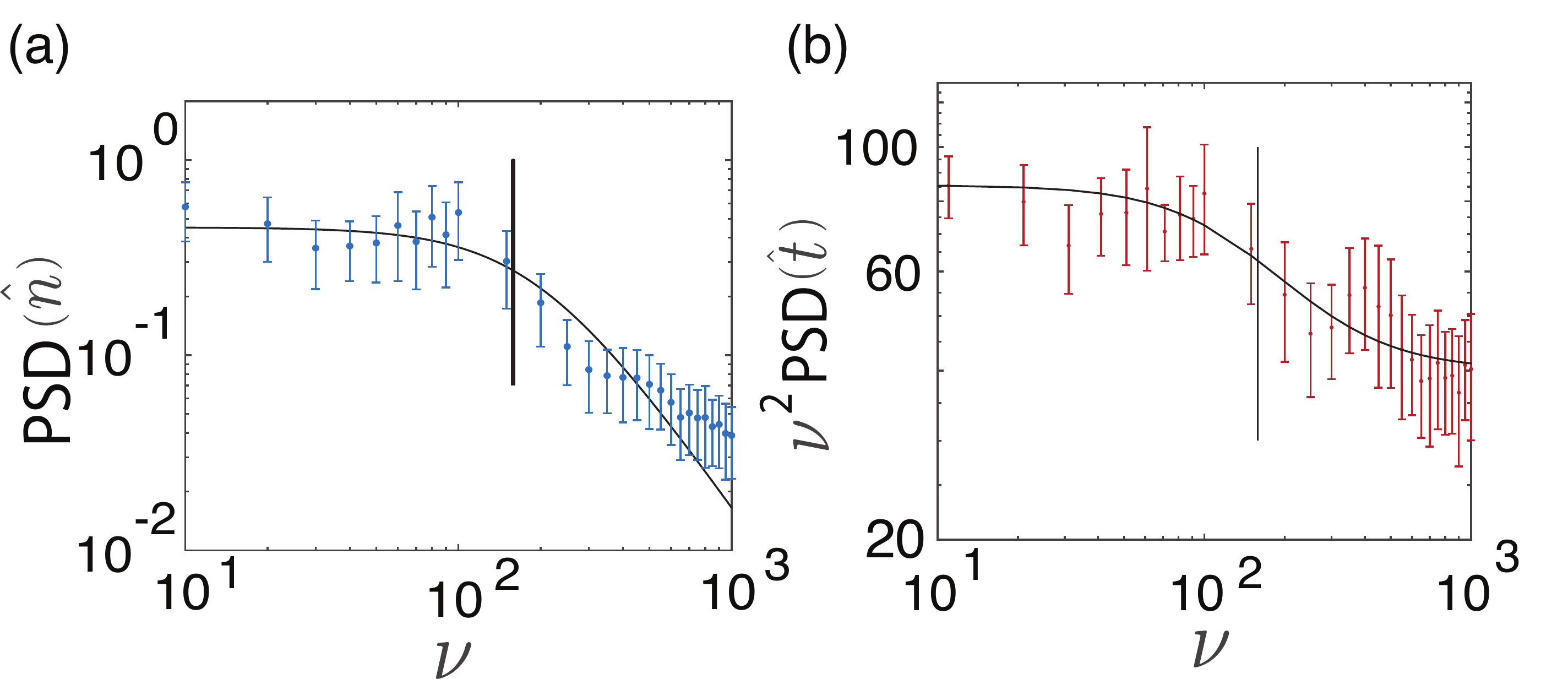}
		\colorcaption{{\bf Experimental fluctuation spectra:} (a) Power spectrum of fluctuations along the local normal to the mean limit cycle, showing a Lorentzian structure. The 
			corner frequency is marked by the vertical black line. Error bars are given by the standard deviation of the mean in each frequency bin.  (b) The observed phase 
			diffusion constant showing the {\em mechanism I} crossover at the corner frequency found from the spectrum in part (a).
			\label{fig:PSDs-hopf-expt1}}
	\end{figure}

	We next compared these limit cycle oscillations to those obtained from the normal-form equation for the Hopf system (Fig.~\ref{fig:Limit-cycle-sim}). While there are geometric differences in the shape of the experimentally obtained mean limit cycle and that predicted by the Hopf normal form, the power spectra of perturbations, computed along the two directions of the Frenet frames, exhibit similar characteristics, as seen from a comparison of Fig.~\ref{fig:PSDs-sim} and Fig.~\ref{fig:PSDs-hopf-expt1}. 

	In Fig.~\ref{fig:PSDs-hopf-expt1}, we observe the expected Lorentzian nature of the perturbations orthogonal to the curve (along $ \hat{n}$) and the 
	diffusive power spectrum for phase fluctuations.  We find the predicted {\em mechanism I} effect in which the observed phase diffusion constant decreases from a 
	larger low-frequency value to a smaller high-frequency one. Moreover, the transition occurs at the corner frequency of the Lorentzian fluctuations, corresponding to fluctuations in the direction normal to the mean limit cycle.  Hence, the fluctuations of the bundle about its mean limit cycle are in agreement with the prediction of 
	the simple supercritical Hopf system. The study of small deviations supports the applicability of the Hopf oscillator description of hair bundle dynamics, 
	even for the bundles situated deeply within the oscillatory regime.

	\subsection{Comparison to the three-dimensional model}
	For direct comparison to predictions based on the more detailed three-dimensional model of the hair cell, we combined measurements of hair bundle displacement with electrophysiological records of the membrane potential. Following techniques previously developed in the laboratory, we patch-clamped the hair cells, under two-compartment configuration, which maintains ionic conditions comparable to those found \textit{in vivo}. These recording conditions maintain the innate bundle oscillations, while providing access to the electrical state of the cell.  These measurements were made in current-clamp mode, yielding data on the time-varying membrane potential ~\cite{Meenderink2015}. The time delay in the voltage recording, due to the pipette resistance and the hair cell capacitance was on the order of $\sim$1 - 2 ms. We neglect this in comparison to the bundle's time period $\sim$30 ms.
	
	As mentioned above, there is to date
	no method of directly accessing internal myosin motor activity.  Thus, we compare our experimental findings to the two-dimensional ${\cal E}$ plane, which constitutes a projection of the full three-dimensional limit cycle, shown in Fig.~\ref{fig:3d-proj-to-2d}. We further note that fluctuations of this system will be studied as deviations from the 
	mean limit cycle, as the ``zero-temperature" limit cycle is not experimentally accessible. 
	
\begin{figure}
		\includegraphics[width=1\linewidth]{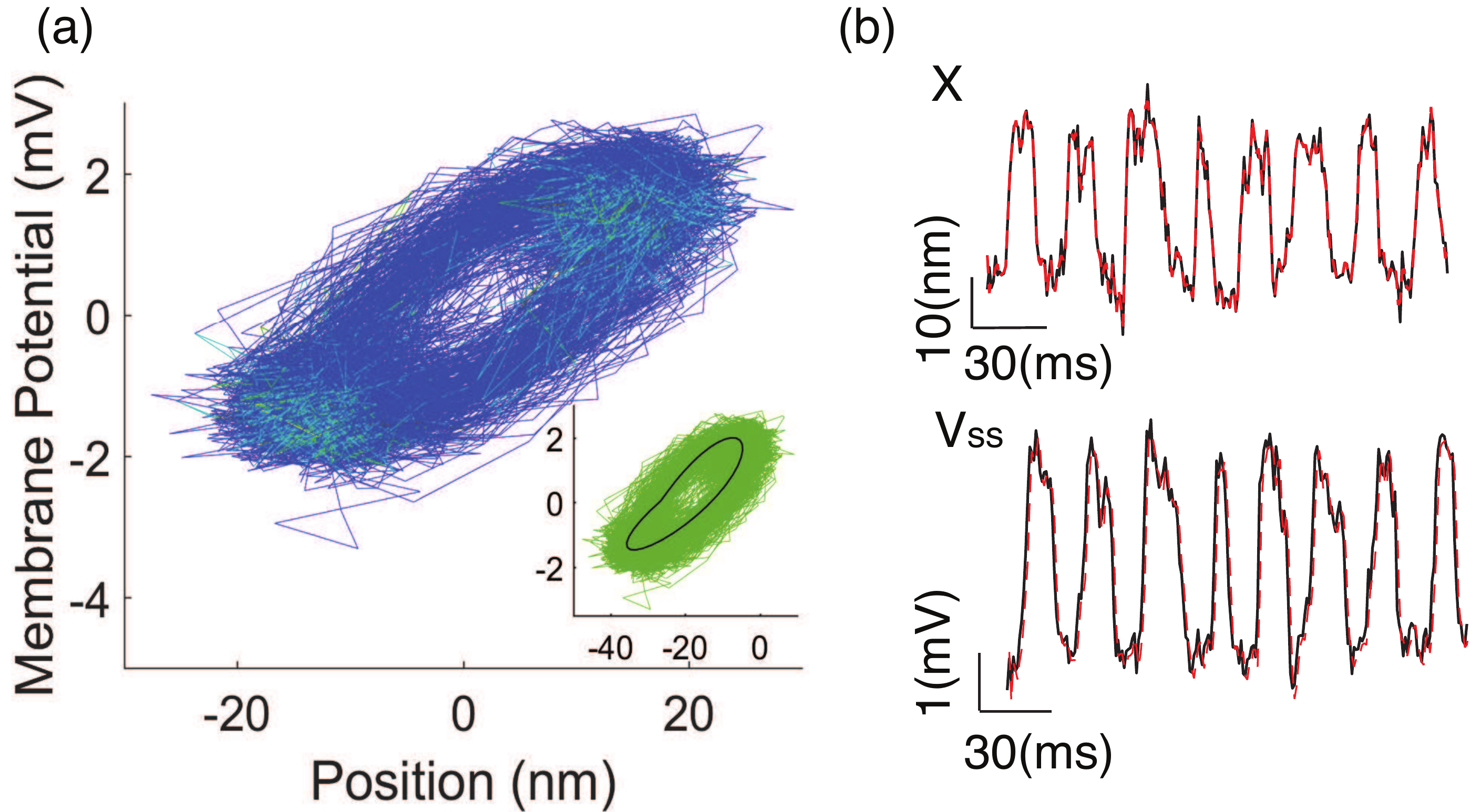}
		\colorcaption{{\bf Experimental observations of bundle position and membrane potential:} (a) Density plot of limit cycle trajectory obtained from simultaneous experimental measurements of the bundle position and somatic potential (blue curve). The light blue regions are more densely populated than the dark blue ones. (Inset) The mean limit cycle is superposed as the black curve. (b) Bundle position $X(t)$ and membrane potential $V_{\rm ss}(t)$ are shown in black. Their low-pass filtered versions are shown as red dashed lines. The resting potential of the hair cell is -67.5 mV, its capacitance is 13.5 pF and the holding current for the current clamp is -2.6 nA.
			\label{fig:limit-cycle-3d-expt1}}
	\end{figure}

	In the inset of Fig.~\ref{fig:limit-cycle-3d-expt1}, we show a typical trajectory of the system in ${\cal E}$ spanned by the state variables $(V_{\rm ss}, X)$ (green curve). Superposed is the mean limit cycle indicated with a black line. In the density plot, similar to Fig.~\ref{fig:velocity-of-limitcycle}, we notice the non-uniformity of the phase space density, thus indicative of the presence of {\em mechanism II}. Panel (b) shows sample traces of the recorded bundle position and somatic potential. 
	
\begin{figure}
		\includegraphics[width=1\linewidth]{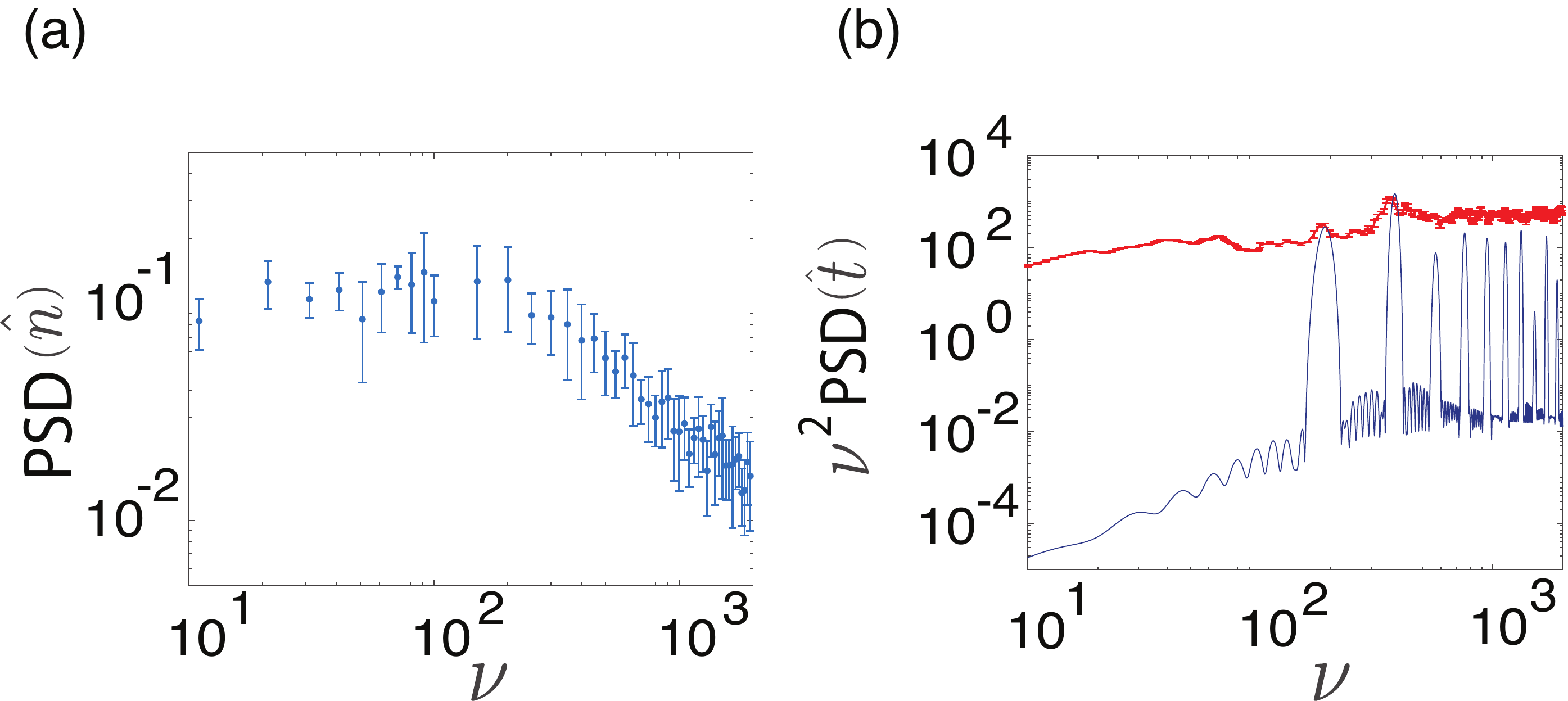}
		\colorcaption{{\bf Experimental fluctuation spectra:} (a) Power spectral density of fluctuations along the $\hat{n}$ direction. (b) Frequency-dependent phase 
			diffusion constant (upper red) along with the Fourier transform of the arclength along the mean limit cycle (lower blue).  The ``Bragg peaks'' of the mean limit cycle appear 
			as less distinct features  in the phase diffusion constant, at the natural frequency and its second harmonic, suggesting a {\em mechanism II} effect.
			\label{fig:PSDs-3d-expt1}}
	\end{figure}

	To explore fluctuation spectra in the experimental recordings, we introduce a Frenet frame associated with the averaged limit cycle and resolve the deviations of the stochastic system from 
	that mean along the local normal and tangent vectors, as done previously. Note that in both Fig.~\ref{fig:PSDs-hopf-expt1} and Fig.~\ref{fig:PSDs-3d-expt1}, we study the system using the Frenet frame vectors $\{ \hat{t}, \hat{n}\}$. However, these systems differ in the experimental variables that are accessed, as the latter includes an independent measurement of $V_{\rm ss}$. 
	
	In Fig.~\ref{fig:PSDs-3d-expt1}(a),  we observe a Lorentzian power spectrum of the fluctuations along $\hat{n}$, consistent with predictions of the numerical model. We plot the frequency-dependent phase diffusion constant in Fig.~\ref{fig:PSDs-3d-expt1}(b), in red.  The shapes of the spectra are comparable to those observed with the $\cal{E}$ projection of the theoretical three-dimensional model (Fig.~\ref{fig:temp-dependence-of-D}), obtained at noise levels comparable to real systems (red curve). The figure includes a plot of the power spectrum of the total phase traversed along the mean limit cycle (blue), which exhibits distinct maxima at frequencies corresponding to the expected peaks of the phase coherent system, i.e., at the natural frequency and its harmonics. The superposition shows that the broad, barely distinct, peaks in the phase diffusion constant occur at frequencies corresponding to the first and second peaks. This correspondence suggests that the system remains phase coherent
	over times greater than one period of the limit cycle. We do not observe the {\em mechanism I} coupling as seen in Fig.~\ref{fig:psd-neiman-3d}, leading to an overall reduction of the phase diffusion constant at 
	higher frequencies.  We return to this point in the discussion.
	
\begin{figure}
		\includegraphics[width=1\linewidth]{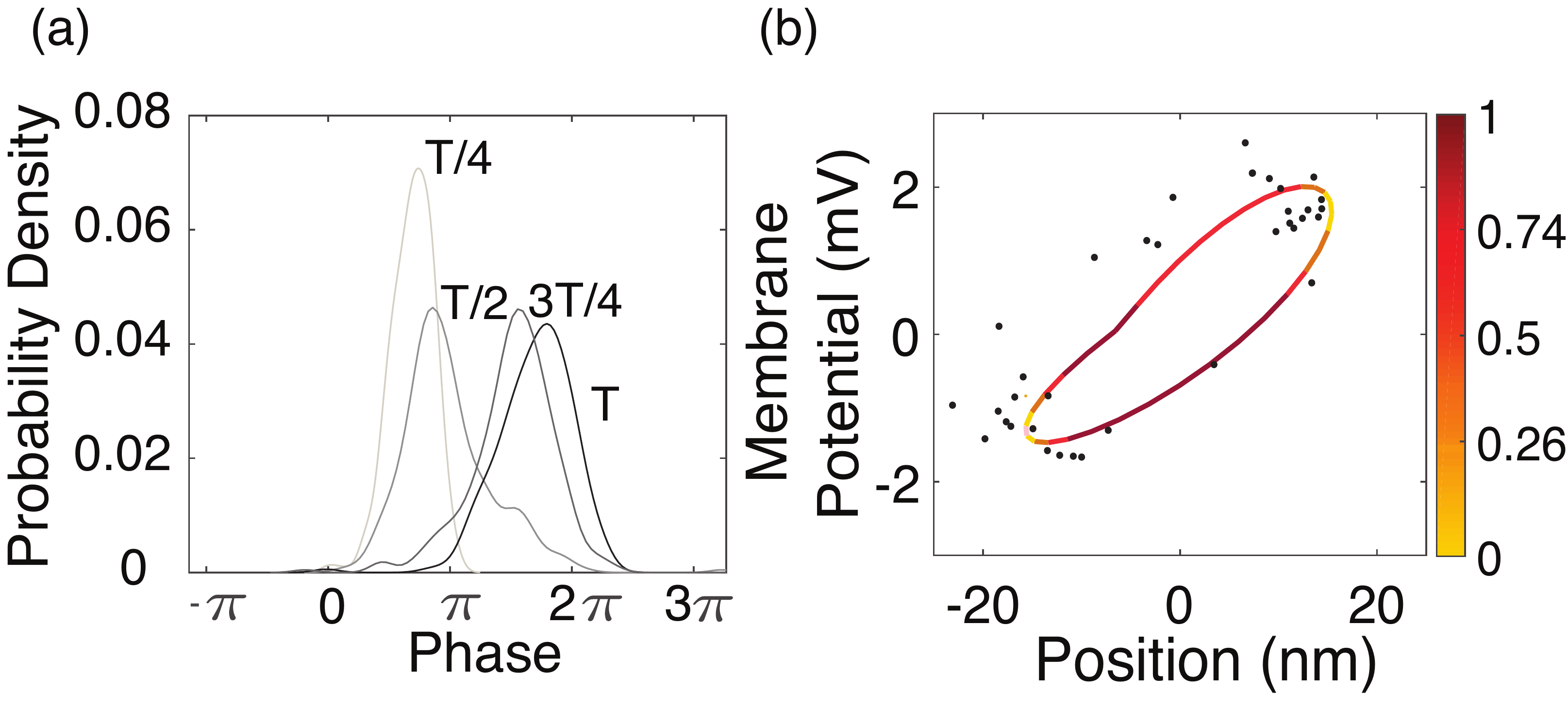}
		\colorcaption{{\bf Measurements of phase diffusion:}  (a) The probability distributions of the phase, after time delays of $T/4, T/2, 3T/4$, and $T$, starting from phase synchronization at an arbitrary point. (b) The experimental mean limit cycle, with the mean phase velocity (normalized to the highest speed) shown as a color map. Redder (darker) colors indicate higher speeds. Data from one cycle of oscillation are shown as black dots. These panels are similar to  Fig.~\ref{fig:phase-distribution}(a) and Fig.~\ref{fig:velocity-of-limitcycle}(b) respectively.
			\label{fig:velocity-limitcycle-expt1}}
	\end{figure}

	We may also consider a number of limit cycle oscillators, all starting at an arbitrary but fixed phase point, to explore the phase advection and diffusion about the limit cycle.  
	The results are shown in Fig.~\ref{fig:velocity-limitcycle-expt1}(a), with phase distributions displayed for times $t = T/4,T/2,3T/4,T$.  As can be seen from the figure, phase coherence persists for at least one limit cycle period, which supports our interpretation of the maxima in the phase diffusion constant at frequencies corresponding to the Bragg peaks. 
	
	We also examine the mean phase velocity around the limit cycle by collecting data from many cycles of the oscillation and binning the observed phase
	velocity by the phase.  The result is shown in Fig.~\ref{fig:velocity-limitcycle-expt1}(b), where the mean phase speed is shown as a color map superposed on an averaged limit cycle, with 
	redder colors representing higher phase speeds.  The series of discrete points (black dots) shows the raw data from one typical trajectory around the limit cycle. 
	This method was tested on six data sets obtained from three cells, each held at two different current-clamp values; similar results were obtained from all of the cells. Another example is presented in Fig.~\ref{fig:velocity-limitcycle-expt2} in Appendix B.
	
	\section{\label{sec:summary}Summary}
	We propose that the study of fluctuations of a stochastic limit cycle oscillator is facilitated by the use of a comoving Frenet frame associated with that limit cycle. 
	This method allows one to resolve the observed fluctuations of a $d$-dimensional system in a natural moving reference frame, which yields a number of advantages. 
	Fluctuations along the local tangent vector and the fluctuations along the $(d-1)$ directions orthogonal to that tangent are predicted to have a simple form, which can be understood in terms of the stability of the limit cycle itself.  Namely, the $(d-1)$ orthogonal directions will behave as (potentially coupled) overdamped oscillators,
	leading to simple Lorentzian fluctuation power spectra.  The phase degree of freedom is meanwhile necessarily diffusive.  
	
	Further, we predict that there are two distinct ways in which fluctuations can couple, leading to more complex spectra.  In {\em mechanism I}, we allow for coupling between the $(d-1)$ normal fluctuations and the phase fluctuation, leading to a crossover from higher to lower phase diffusion constant. The crossover occurs at the corner frequency of the Lorentzian fluctuation spectrum of the orthogonal  degrees of freedom.  In {\em mechanism II}, we find that the confining potential of the orthogonal fluctuations or the mean phase velocity can be phase-dependent. The phase diffusion constant can thus acquire more structure, including local maxima or ``Bragg peaks'' at frequencies corresponding to the mean period of the oscillator.  We verified these ideas through numerical simulations based on both a simple model, the normal form supercritical Hopf model, and a more complex one, a three-dimensional model describing the specific biophysical processes of the hair cell of the inner ear.  
	
	Applying this analysis to experimental data obtained from oscillating hair cells, we find that the supercritical Hopf model not only accurately predicts the mean limit cycle oscillatory behavior, but 
	also describes well the fluctuation spectra normal to the limit cycle and the phase diffusion about it. In particular, we observe strong evidence for 
	the predicted {\em mechanism I} coupling between the phase and normal fluctuations in the frequency dependence of the phase diffusion constant.  Thus, the 
	simplest supercritical Hopf model accurately accounts for both the mean dynamics, and the small fluctuations about that mean. One implication of this result is that
	the stochastic forces acting on the bundle appear to be adequately described by white noise. If there were just Brownian forces in the surrounding viscous fluid, this 
	point would be unremarkable. However, the total noise in the system must include stochastic effects in various active elements, which may generate colored noise. Our results 
	then constrain the frequency dependence of such stochastic forces acting on the Hopf model of the bundle.
	
	The symmetry of the Hopf model precludes a {\em mechanism II} coupling here. We do not observe it in either the model (Fig.~\ref{fig:PSDs-sim}) or in the experimental data (Fig.~\ref{fig:PSDs-hopf-expt1}). When we examine the more complete data sets, combining cell potential and bundle deflection, we do not observe (in Fig.~\ref{fig:PSDs-3d-expt1})
	significant effects of {\em mechanism I} coupling between the normal and phase variables. Such a coupling is only weakly seen in the full 3d model (Fig.~\ref{fig:psd-neiman-3d}). When we project the full model onto the experimentally observable plane - see Fig.~\ref{fig:temp-dependence-of-D} - the evidence of the {\em mechanism I} coupling vanishes. The projection to the ${\cal E}$ plane appears to mask the frequency structure of the {\em mechanism I} coupling in the phase diffusion. Because of that projection, the tangent and normal vectors to the observable but projected limit cycle are superposition of the true Frenet frame vectors associated with the limit cycle of the full dynamical system. We speculate that this projection onto the experimental manifold ${\cal E}$ also masks the effect of the {\em mechanism I} coupling in the experimental data - see Fig.~\ref{fig:PSDs-3d-expt1}.
	
	A number of other questions about stochastic limit cycle dynamics remain open.  There is, as yet, no general theory for how the average limit cycle of a system 
	changes size and shape as a function of noise amplitude or temperature. The ``thermal expansion'' of the limit cycle and the change in its mean period must result
	from a combination of the asymmetry of the confining potential about the limit cycle and the Kramers escape ``corner-cutting'' mechanism mentioned previously.  We expect the mean limit cycle to distort with temperature in a manner controlled by the nonlinearities of the confining potential, and to observe a smoothing of any sharp folds 
	due to corner-cutting by noise-activated processes.  These ideas will be explored in a future study. 
	
	The results of the comparisons between the theoretical models and the data suggest another avenue for further study.  We expect that {\em mechanism I} effects should allow one 
	to assess how many hidden variables are necessary to fit both the mean limit cycle and the observed fluctuations, especially phase diffusion.  The {\em mechanism I} coupling requires fluctuations along a local normal to the limit cycle. The observation of $n$ steps illustrative of such a coupling then necessitates at least $n$ normal directions. Here, step refers to transitions as depicted in Fig.~\ref{fig:PSDs-sim}(b). Since {\em mechanism I} couplings can be masked or otherwise be rendered unobservable (as seen in the projection of the 3d model to the ${\cal E}$ plane), the number of such observed steps sets a lower bound on the dimensionality of the underlying dynamical model of the hair bundle. The further analysis of the fluctuation spectrum may also point to the presence of colored, nonequilibrium noise in the system. 
	
	This analysis also suggests new experiments designed to probe the changes in the fluctuations and shape of the limit cycle in response to the variation of various model 
	parameters.  The most pertinent would be an analysis of the change in the limit cycle in response to perturbations in myosin motor activity.  Possible avenues include modulation of myosin activity through pharmacological manipulations, specifically, interference with its phosphorylation pathway, or modulations of the calcium concentration.  Other experimental perturbations include variation of the temperature of the biological system, loading the bundle with an elastic element, and interference with the fluctuations of the membrane potential. 
	
	Finally, we note that there is growing interest in understanding the thermodynamics of nonequilibrium steady states. For example, there have been proposed generalization of the fluctuation-dissipation theorem ~\cite{Chaikin1995} to nonequilibrium, but time-independent states. Considerations of out-of-equilibrium systems near a stationary fixed point have led to generalized fluctuation-dissipation  theorems (GFDTs) ~\cite{Prost2009,Dinis2012}.  Further theoretical work includes the extension of these generalizations to limit cycle oscillators; as these are periodic in time, we do not expect a generalization of the FDT to involve only one frequency. Nonetheless, we expect that there may well be a relation between a suitably defined two-frequency correlation function of one or more of the dynamical variables and their time (and phase of the 
	limit cycle) dependent response to conjugate forces.  
	\\

\acknowledgements

The authors would like to thank Justin Faber, Tracy Zhang, and Elizabeth Mills, for providing recordings of spontaneous bundle oscillations analyzed in this paper.

DB acknowledges support from NSF PoLS grant 1705139, and NIDCD grant R21DC015035. AJL acknowledges partial support from NSF-CMMI-1300514 and NSF-DMR-1709785. 

%

\appendix
\section{Simulation details}
The stochastic simulations of Eq.~\ref{Hopf-main} were carried out using the 4$^{th}$-order Runge-Kutta 
method for a duration of 60 s. The corresponding time steps were in the range of $10^{-4}  \leftrightarrow 2 \times 10^{-3}$ s.   The experimental data were obtained with
time steps of  $10^{-3}$ and $ 2 \times 10^{-3}$ s.  

Experimentally, the variance of the noise experienced by the hair bundle, normalized by the square of its drag coefficient ($\frac{2kT}{\lambda}$), is of the order of $3 \times 10^{-6} \frac{nm^2}{s}$, while amplitudes of spontaneous bundle oscillation typically vary from 10 -- 50 nm. The noise variances $A_{ij}$ in the Hopf simulations were varied from 
$10^{-7} \leftrightarrow  0.4$, with bundle oscillation amplitude fixed at $1$; consistent results were obtained over the full span of noise amplitudes.  Fig.~\ref{fig:Limit-cycle-sim} employs the highest variance value in this range. The stochastic terms driving $\left\{ x(t), y(t) \right\}$ were assumed to be uncorrelated.  This assumption may be relaxed in future work. 

Eqs. ~\ref{Neiman-model-pos}, ~\ref{Neiman-model-myosin}, and ~\ref{Neiman-model-V} were integrated 
using the Euler-mayurama method for a duration of 60 s using time steps of  $5 \times 10^{-4}  \leftrightarrow 10^{-3}$ s. The noise statistics of the bundle and myosin motors were assumed to follow white Gaussian ensembles with zero means 
and variances of $\left\{2kT\lambda,  2kT*1.5\lambda_a\right\}$, respectively~\cite{Nadrowski2004}. 
The values of the parameters used in the simulations are given in Table~\ref{table:parameters}.
\begin{table*}
	\begin{threeparttable}
\caption{Model Parameters}
\centering
\begin{tabular}{l c c}
\hline
\hline
Symbol & Values & Parameter  \\ [0.5ex]
\hline
K$_{gs}$ & 750$\frac{\mu N}{m}$ & gating spring constant \\
D & 62.1$nm$ & gating compliance \\
K$_{sp}$ & 600$\frac{\mu N}{m}$ & stereociliary pivot spring  \\
$\gamma$ & 0.14 & geometric coefficient \\
F$_{\rm max}$ & 500 pN & maximal force exerted by adaptation motors \\
P$_0$ & 0.63 & probability of channel opening \\
$\alpha$ & 0.8 & Ca$^{2+}$ feedback on motors \\ 
$\omega_v$ & 2$\pi \times 20$ &  frequency of voltage oscillations, in the absence of the MET current \\
Q$_v$ & 30 & quality factor in the absence of MET current \\
C$_m$ & 14 pF & capacitance of a hair cell \\
$\beta_0$ & $\frac{\omega_v C_m}{Q_v}$ &  damping coefficient in the absence of MET current \\
g$_t$ & 1.5 nS & conductance of transduction channels \\
V$_0$ & -55 mV & resting potential of the cell \\
I$_0$ & 10 pA & leakage current  \\
$\lambda$ & 2.8 $\frac{\mu Ns}{m}$ & bundle drag coefficient \\
$\lambda_a$ & 10 $\frac{\mu Ns}{m}$ & motors drag coefficient \\
$k_{B}$ & 1.38 $\times$ 10$^{-23}$ $\frac{m^2 kg}{s^2 K}$ & Boltzmann constant \\
$T$ & 300 $K$ & Room temperature \\
\hline
\end{tabular}
\label{table:parameters}

 \begin{tablenotes}

	\item These parameter values were obtained from references \cite{Nadrowski2004} and \cite{Neiman2011}.
\end{tablenotes}
\end{threeparttable}
\end{table*}
All simulations were performed in MATLAB (R2017a, the MathWorks, Natick, MA).

\section{Experimental data}

\begin{figure}
\includegraphics[width=1\linewidth]{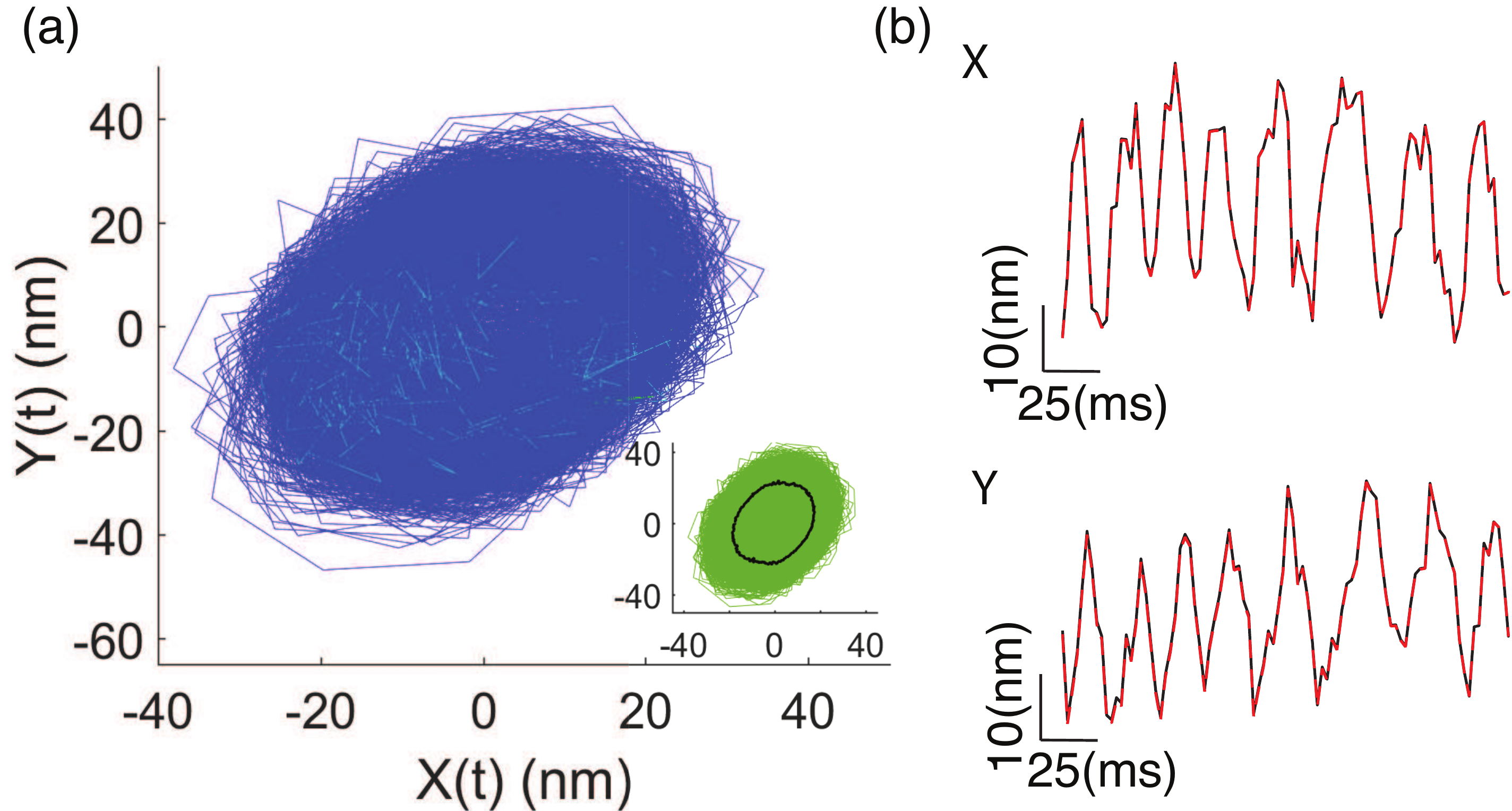}
\colorcaption{{\bf Experimental recording of an irregular oscillator:} (Left) Typical trajectory (green) and averaged limit cycle (black) for a cell showing less well defined limit cycle dynamics. (Right) Time 
series of the bundle position and velocity (obtained via a Hilbert transform) after low-pass filtering.
\label{fig:limit-cycle-expt2}}
\end{figure}

Significant cell-to-cell variation was observed in both the mean limit cycles exhibited by active hair bundles and in the fluctuations about those limit cycles.  In the main 
text, we show a representative data set for a cell that showed relatively regular oscillations.  A number of cells showed less regular limit cycles; we show here 
a data set representative of this type of cell.  In this case, rather than forming a ring, the trajectories about the limit cycle appear to fill a disk, as shown in Fig.~\ref{fig:limit-cycle-expt2}.  

\begin{figure}
	\includegraphics[width=1\linewidth]{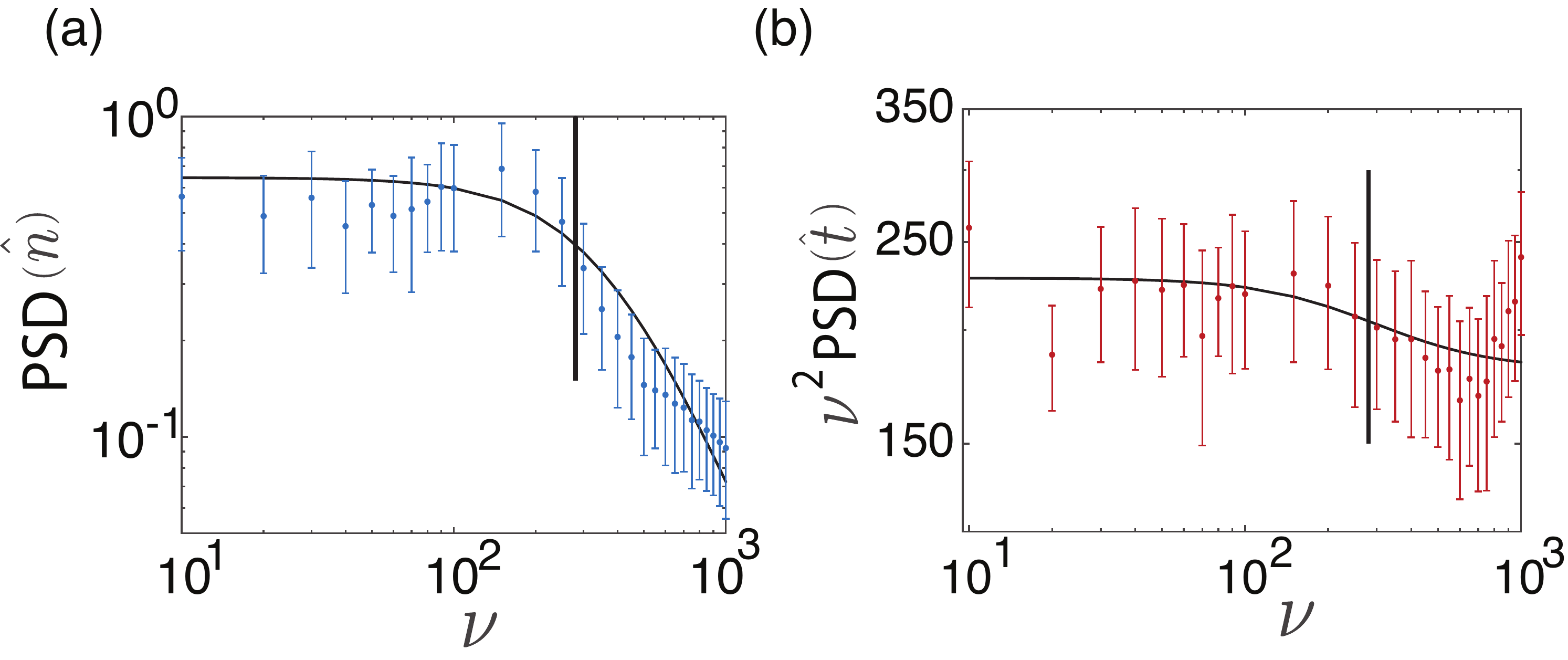}
	\colorcaption{{\bf Fluctuation spectra of an irregular oscillator:}  (A) Power spectrum of the fluctuations in the normal direction (blue), along with a best fit Lorentzian (black). (B) 
		Frequency-dependent phase diffusion constant. In both 
		panels, the vertical (black) line indicates the corner frequency.
	\label{fig:PSDs-hopf-expt2}}
\end{figure}

Despite the more noisy limit cycle dynamics, one can extract the fluctuations along the local normal $\hat{n}$ and local tangent $\hat{t}$ to the 
averaged limit cycle.  The power spectra of the fluctuations are shown in Fig.~\ref{fig:PSDs-hopf-expt2}.  The corresponding data set, is filtered with a cut-off at $225\times2\pi$ Hz; the Nyquist frequency is $250\times2\pi$ Hz. We note that the fluctuations in the normal direction (blue) are still well described by 
a simple Lorentzian (black) and obtain a corner frequency from this fit, which is denoted by the black vertical line.  The phase diffusion constant shows, however, only a weak frequency dependence.  The {\em mechanism I} transition from a larger to a smaller diffusion constant at the corner frequency is, at best, suggested
by these data.

\begin{figure}
	\includegraphics[width=1\linewidth]{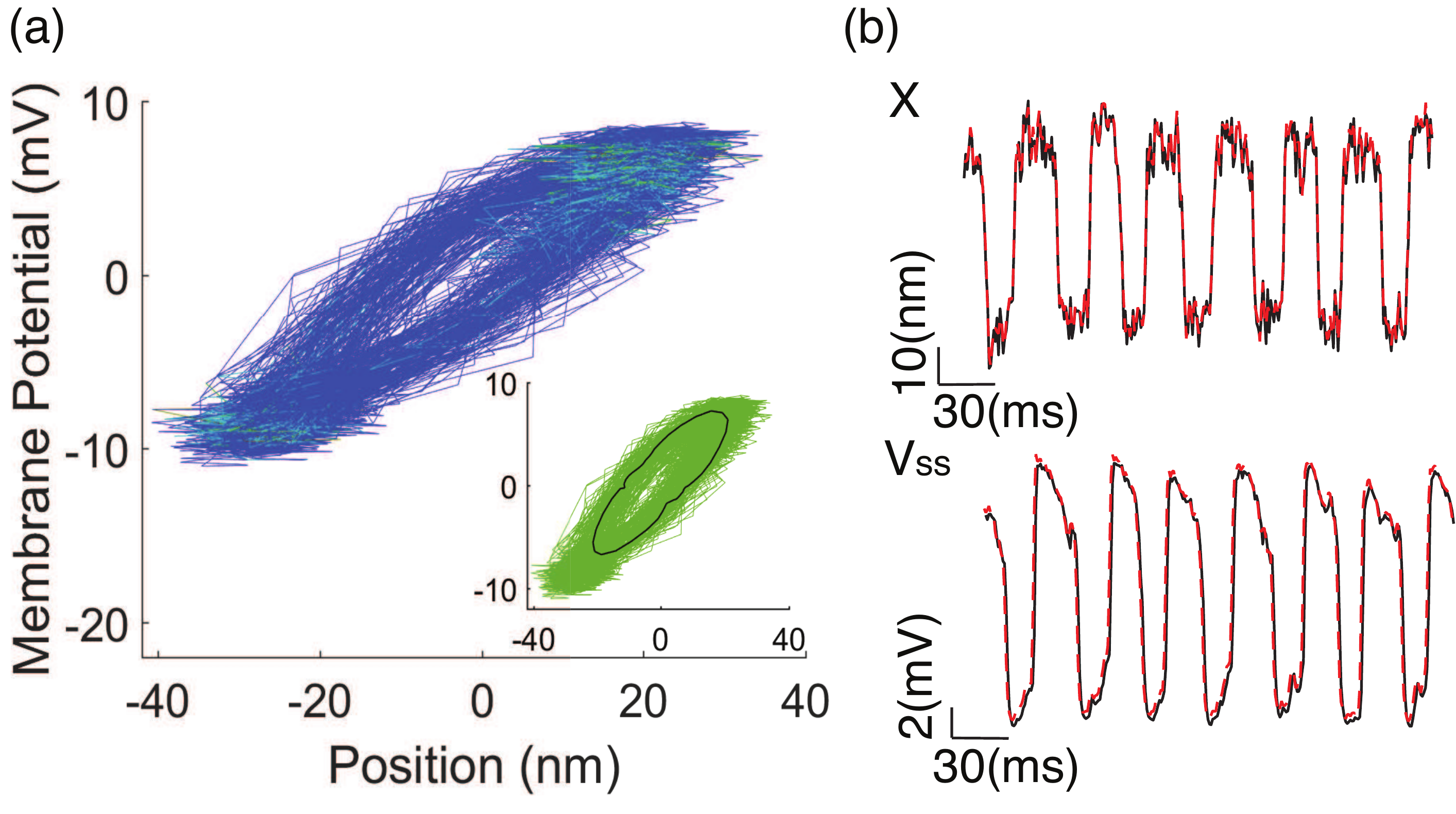}
	\colorcaption{{\bf Experimental recordings of a noisy hair cell:} (Left) Stochastic trajectory (green) with the averaged limit cycle (black) in the space spanned by bundle deflection $X$ and membrane
	potential $V_{\rm ss}$.   (Right) Time series of these dynamical variables, without (black) and with (red) low-pass filtering. The resting potential of the hair cell is -31 mV and the holding current -1.5 nA.
	\label{fig:limit-cycle-3d-expt2}}
\end{figure}

\begin{figure}
	\includegraphics[width=1\linewidth]{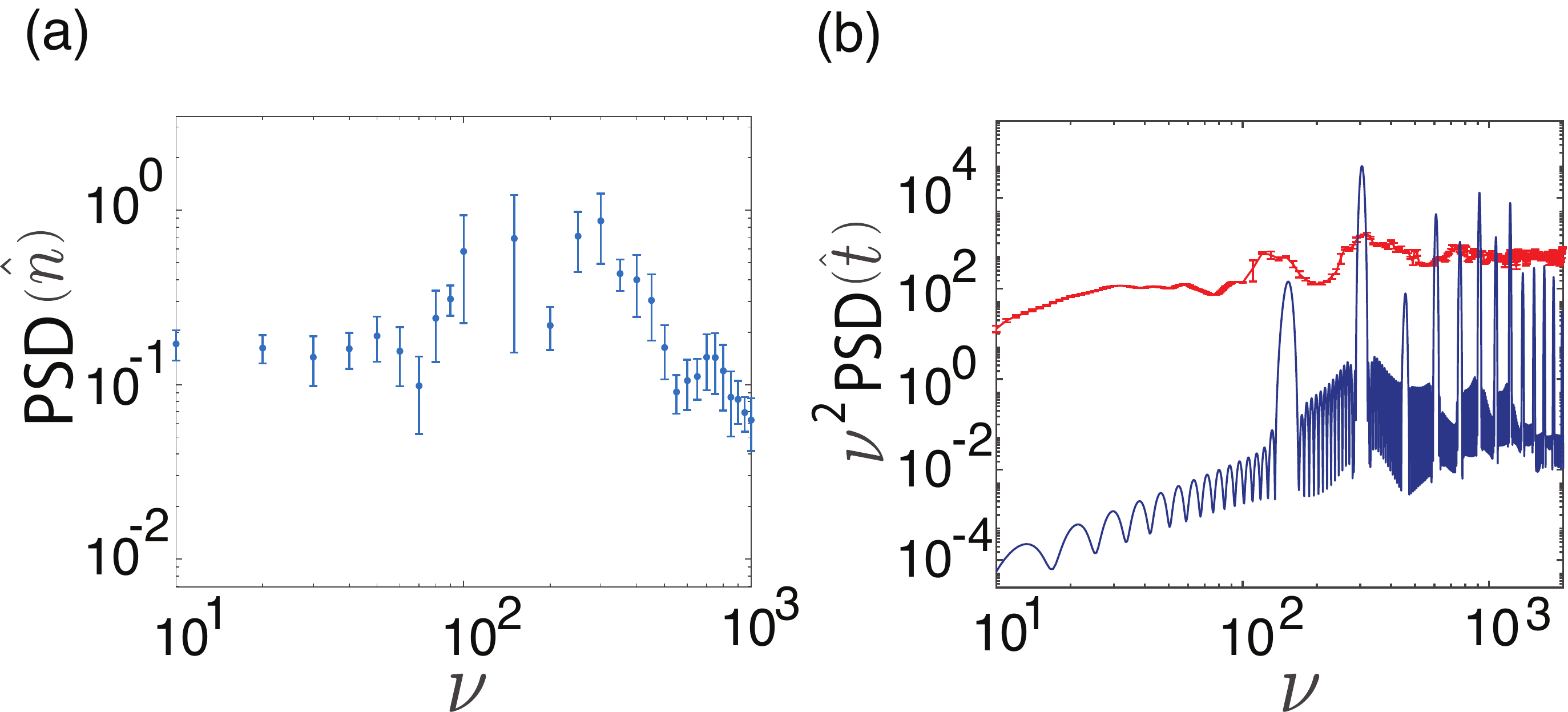}
	\colorcaption{{\bf Experimental fluctuation spectra:} (A) Power spectrum of fluctuations normal to the limit cycle. (B) Phase diffusion constant, 
	 exhibiting broad peaks at the natural frequency of the bundle and its second harmonic. The power spectrum (red) is qualitatively similar to that observed in simulations with  
	 noise variance of $\tau_{\rm noise} / \tau_{0} = 1$.  The power spectrum (dark blue) of the total phase traversed by the system along the average limit cycle is shown to locate the 
	 expected peaks.
	\label{fig:PSDs-3d-expt2}}
\end{figure}

\begin{figure}	\includegraphics[width=1\linewidth]{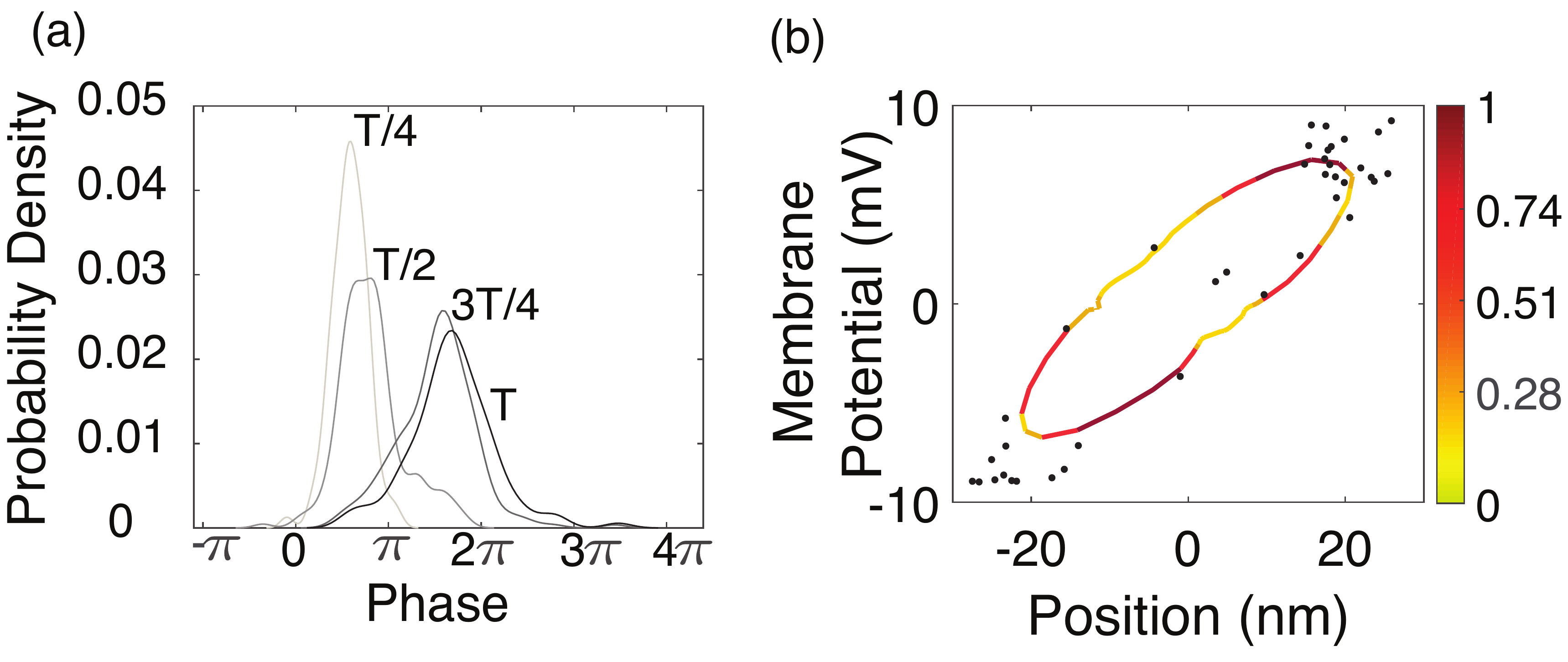}
	\colorcaption{{\bf Experiment:}  (A) The probability densities of initially phase synchronized ensembles, illustrating phase diffusion. (B) Mean limit cycle velocities shown as a color map on the mean limit cycle with (dimensionless) velocity increasing from colder (blue) to  
		warmer colors.
	\label{fig:velocity-limitcycle-expt2}}
\end{figure}

We find similarly large cell-to-cell variations in the dynamics of hair cells when we obtain both bundle position and membrane potential.  
Another example of a bundle (green) described experimentally by its position and membrane potential is shown in Fig.~\ref{fig:limit-cycle-3d-expt2}, along with its averaged limit cycle (black). 
 Once again, we obtain similar power spectra for fluctuations 
in directions normal and tangent to this averaged limit cycle. The spectra are shown in Fig.~\ref{fig:PSDs-3d-expt2}.  The phase diffusion dynamics obtained from the noisier cell were comparable to those shown in the main text - see Fig. ~\ref{fig:velocity-limitcycle-expt2}. 

\end{document}